\documentclass[12pt]{article}
\usepackage{graphicx}

\newcommand\D{\displaystyle}

\newcommand\be{\begin{equation}}
\newcommand\ee{\end{equation}}
\newcommand\ba{\begin{eqnarray}}
\newcommand\ea{\end{eqnarray}}
\newcommand\bano{\begin{eqnarray*}}
\newcommand\eano{\end{eqnarray*}}

\def\NP#1{Nucl. Phys.~{\bf #1}}
\def\NPPS#1{Nucl. Phys. Proc. Suppl.~{\bf #1}}

\def\PL#1{Phys. Lett.~{\bf #1}}
\def\PR#1{Phys. Rev.~{\bf #1}}
\def\PRP#1{Phys. Rep.~{\bf #1}}

\def\ap#1{Annals of Physics {\bf #1}}

\def\npa#1{Nucl. Phys. {\bf A#1}}
\def\npb#1{Nucl. Phys. {\bf B#1}}

\def\plb#1{Phys. Lett. {\bf #1B}}

\def\prd#1{Phys. Rev. {\bf D#1 }}
\def\prle#1{Phys. Rev. Lett. {\bf #1}}
\def\ptp#1{Progr. Theor. Phys. {\bf #1}}

\def\zpc#1{Z. Phys. {\bf C#1}}

\begin{document}
\setcounter{page}{0}
\def\footnoterule{\kern-3pt \hrule width\hsize \kern3pt}
\title{\vskip -2cm\hfill{\small IKDA 99/04}
\vskip 1cm
Renormalization Group Flow\\ and\\
  Equation of State of Quarks and Mesons}

\author{
	Bernd--Jochen 
	Schaefer \thanks{Email: {\tt Schaefer@NHC.tu-darmstadt.de}}\\[3mm] 
    {\em Institut f\"ur Kernphysik}\\
    {\em Technische Universit\"at Darmstadt}\\
    {\em D--64289 Darmstadt}\\[6mm]
	and\\ Hans--J\"urgen 
	Pirner \thanks{Email: {\tt pir@dxnhd1.mpi-hd.mpg.de}}\\[3mm]
    	{\em Institut f\"ur Theoretische Physik} \\ 
	{\em Universit\"at Heidelberg} \\ 
	{\em Philosophenweg 19} \\ 
	{\em 69120 Heidelberg, Germany}
	}

\maketitle

\thispagestyle{empty}

\begin{abstract}
Nonperturbative flow equations within an ef\-fect\-ive
linear sigma model coupled to constituent quarks 
for two quark flavors are derived and solved. 
A heat kernel regularization is employed 
for a renormalization group improved effective potential.
We determine the initial values of the coupling constants in
the effective potential 
at zero temperature. 
Solving the evolution equations with the same initial values
at finite temperature in the chiral limit, 
we find a second order phase transition 
at $T_c \approx 150$ MeV. 
Due to the smooth decoupling of massive modes, we can directly link
the low-temperature four-dimensional theory to the three-dimensional
high-temperature theory. We calculate the equation of state 
in the chiral limit and for finite pion masses and determine 
universal critical exponents.  
\end{abstract}

\vspace*{\fill}

\section{Introduction}
At zero temperature and zero chemical potential, the chiral symmetry
of Quantum Chromodynamics (QCD) is spontaneously broken. One expects that the
chiral symmetry will be restored at sufficiently high
temperature and a phase transition occurs 
separating the low temperature and high
temperature regions. This phenomenon may be realized in high energy heavy-ion
collisions at RHIC (Brookhaven) and LHC (Geneva) 
and is therefore a subject of general
interest. It is a matter of intense debate at what temperatures 
the phase transition would occur and what its nature is.

In this paper we study the chiral phase transition at finite
temperature both  in the chiral limit and for
finite pion masses.
We present flow equations which link 
the universal critical behaviour at finite temperature to the well
known physics at $T=0$. Calculating the effective potential in the
framework of the renormalization group theory with a
heat kernel regularization, we
investigate the critical behaviour of the chiral phase transition for
two light quark flavors  and 
calculate the critical exponents. It is still an open question whether 
full QCD with two massless 
quarks\footnote{The influence of the
	strange quark is neglected in this work. 
	}
and the three-dimensional $O(4)$
Heisenberg model 
belong to the same universality class.  
We use the linear sigma model with quarks which
at $T=0$ exhibits a spontaneous breaking of the $O(4)$ symmetry to 
an $O(3)$ symmetry. We show how the $O(4)$ symmetry  
is restored at sufficiently high
temperature.
Lattice investigations hint that the two flavor chiral phase
transition lies  in the $O(4)$
universality class \cite{kana} from which the behaviour of the
condensates and the long distance properties at the critical
temperature ensue. If this is confirmed  the transition would be 
basically driven 
by the pions and the chiral sigma particle.

It was realized that a perturbative expansion of the
effective potential breaks down near the critical temperature due to
infrared divergences and the nature of the phase transition cannot
be investigated in any finite order in perturbation theory. The
situation was partly resolved in \cite{wett} by means of  
the average potential which employs renormalization group ideas. 
Constructing evolution equations for the dependence of the effective
potential on a variable infrared cutoff this approach sums
effectively 
the relevant class of perturbative diagrams
in the complicated infrared regime.
We follow this method and calculate in a
nonperturbative way the effective potential 
using a variable heat kernel cutoff. 
The advantage of this cutoff method is that the evolution equations 
can be obtained in analytical form which makes the physics content
very transparent. The aim of this paper is twofold: We demonstrate how
the renormalization group improvement of the mass and coupling constant 
evolution 
give the main physics of the chiral phase transition, i.~e.~the equation
of state away from the critical temperature.
At low temperatures one sees the success of the model  handling the pion
and sigma degrees of freedom in a comparable accuracy to chiral 
perturbation theory. At intermediate temperatures the quarks take over as 
main agents contributing significantly to the change of the chiral
condensate. 
The high temperature behavior of the linear sigma model 
has not enough degrees of freedom to model full QCD lacking in particular
gluons. 

By explicitly constructing the thermodynamic equation of state we think 
a more realistic assessment of the contents of the model is possible.
We also think that our method is very pedagogical making 
this powerful method available to physicists
who are interested in an intuitive access to an otherwise very technical
subject.
 
The paper is organized as follows: In section \ref{sec2} a general
summary of our formalism for an $N$-component $\phi^4$ theory coupled
to $N_f$ quarks is given. After the calculation 
of the effective potential our ansatz for  
renormalization group improvement is 
presented. In section \ref{secchirallimit} 
the nature of the chiral phase transition is elucidated.
The critical behaviour is discussed and  critical exponents are 
determined. In this section 
the free energy density i.e.
the thermodynamic potential for the transition and 
the equation of state 
is calculated in the chiral limit.
Section \ref{secfinitepionmass} 
is reserved for a discussion of the same properties
for finite pion mass.
Section \ref{sec5} presents our conclusions.
In order to improve the readability of this paper we
postpone extensive equations to  appendices.


\section{Flow Equations}
\label{sec2}
Wilson's original renormalization group ideas have been used 
widely in perturbative and nonperturbative
calculations. A nonperturbative approach has been used  on  lattices 
and also implemented in continuum Quantum Field Theory by Wegner,
Houghton and Polchinski 
\cite{wils},\cite{wegn}.  
Flow equations describe the average of an effective action over a
volume and are the continuum  analogue 
of a block spin transformation
on the lattice.
All degrees of freedom with momenta larger than an
infrared scale $k$, the coarse graining scale, are effectively
integrated out. The flow equations are ultraviolet and infrared
finite through the introduction of an infrared cutoff function $f_k$,
which obeys several requirements described in the next subsection. The
solution of these equations 
provides the full effective potential, once the infrared
cutoff $k$ is removed ($k \to 0$). 
On the other side in the limit $k \to \infty$ or $k$
equal to some ultraviolet cutoff $\Lambda$ the effective average
potential results in the  classical  potential where
no fluctuations are included. 
This ultraviolet cutoff
may  be associated with the highest momentum scale
for   which the
theory parameterizes  the physics adequately.
All quantum fluctuations with momenta   larger than $\Lambda$ are
included in the potential at this scale, thus no further
necessity for an ultraviolet regularization scheme arises. The 
effective couplings at the large scale $\Lambda$
specify 
the potential completely. 
The knowledge of the $k$-dependence of the effective average potential
then allows to interpolate from the classical 'bare' potential at $k =
\Lambda$ down to the effective 'renormalized' 
potential for $k \to 0$ including more
and more quantum fluctuations as the evolution proceeds towards zero.

This method of the flow equation has been developed intensively by 
Wetterich and his collaborators \cite{wett}.
One important ingredient of the evolution equation approach by
this group  is  
the appearance of the exact inverse propagator in the flow equation,
which makes the 
RG-improved one-loop equation exact and intrinsically 
complicated \cite{wet1}. The
solved truncation of this exact equation is very similar to the 
RG-improved one loop equations used here where 
the bare couplings are replaced by   
their corresponding full $k$-dependent expressions. 
The technically difficult part in
the exact RG-approach is not the establishment of an exact relation, 
but rather the choice of
a suitable nonperturbative truncation scheme
in order to solve this highly nonlinear coupled set of differential
equations. 
We use a simple truncation scheme and analytical 
flow equations can be derived.
We believe that a deeper connection between Wetterich's 
and our approach should be possible.

The flow equations are nonperturbative, i.e. also
applicable for problems with large couplings,
loosely spoken, because the evolution of the different couplings
is interlinked and no particular truncation in 
coupling constant is used. The  
solutions of the flow equations give an accurate 
picture of the high temperature phase transition in four
dimensions. It is possible to study the phase transition at the 
critical temperature and
to calculate the critical
exponents with  surprisingly high accuracy. 


\subsection{The Effective Potential}

In this subsection we develop the formalism starting with 
the derivation of the flow equations for the linear 
$SU(2)\times SU(2)$ sigma model
combined with two quark flavours. The linear sigma model is
a  hybrid model which
contains quark and mesonic $\sigma$-- and $\pi$--fields as collective 
degrees of freedom in order to 
simplify  the 
strong $q \bar q$ interaction.  
The cutoff $\Lambda$ plays the role
of a compositeness scale, below which the effective $\sigma$- and 
$\pi$-fields can be treated as elementary.
For the moment we leave the connection of this theory to
QCD as an open question and see how far we get with this model.
Preliminary investigations in this direction have been made in 
ref.~\cite{ellw}.

The partition function at zero
temperature is given by 
\ba
Z[ J=0 ] & = & \int {\cal D}q{\cal D}\bar{q} {\cal D}\sigma {\cal D}\vec{\pi}
\exp \{- \int d^4 x ( {\cal L}_F + {\cal L}_B ) \}\quad.
\ea
Here we omit any external sources $J$, because in the one loop 
approximation the effective potential and the generating functional for 
one particle irreducible Green functions do not differ.
The Euclidean\footnote{We denote all Euclidean operators with an index 
  'E'.}  Lagrangian in $d=4$ dimensions looks like
\ba
{\cal L}_F & = & \bar{q}(x) \left( \gamma_E \partial_E +
g\left( \sigma + i\vec{\tau}\vec{\pi}\gamma_5 \right)\right) q(x)\quad,\\
{\cal L}_B & = & \frac{1}{2} \left(
(\partial_\mu \sigma)^2 + (\partial_\mu \vec{\pi})^2 \right) +
\frac{m_0^2}{2}(\sigma^2+\vec{\pi}^2)
+\frac{\lambda_0}{4}(\sigma^2+\vec{\pi}^2)^2 -c \sigma \quad.
\ea
The parameter $c$ is an explicit symmetry breaking term in
$\sigma$-direction,  which 
gives the Goldstone boson a
finite mass. 
In order to avoid later any recurrences we derive the effective 
potential with an explicit symmetry breaking term $c$. 
The chiral limit can be extracted by putting $c$ to 
zero in all equations.

Taking advantage of the properties of
the minimum  $\phi_0$ of the potential  we can eliminate 
the mass parameter $m_0^2$ 
in favour of $\lambda_0$
and $\phi_0$.
Then the potential can
be written with $m_0^2 = c/\phi_0 -\lambda_0 \phi^2_0$ as
\be
 V = \frac{\lambda_0}{4} \left( \vec{\phi}^2 - \phi_0^2 \right)^2
- \frac{\lambda_0}{4} \phi_0^4 -c \left( \sigma - 
\frac{\vec{\phi}^2}{2\phi_0 }\right)
\ee
where we  introduce the $O(4)$ vector $\vec{\phi} = (\sigma, \vec{\pi})$
as a shorthand notation for the meson fields.

Formal integration over the fermions yields a non-local determinant, which
can be regularized by the heat kernel representation. 
We use the abbreviations 
$M(x) = \sigma (x) + i\vec{\tau}\vec{\pi} (x) \gamma_5$, 
($MM^+ = \vec{\phi}^2$)  and
$D = \gamma_E \partial_E + g M(x)$.
\ba
Z[ J=0 ] & = & \int  {\cal D}\sigma {\cal D}\vec{\pi}
\det \left(\gamma_E \partial_E + g M(x) \right)
\exp \{- \int d^4 x  {\cal L}_B  \}\nonumber\\
& = & \int  {\cal D}\sigma {\cal D}\vec{\pi}
\exp \{ \frac{1}{2} Tr \log DD^+ -\int d^4 x  {\cal L}_B  \}\\
& = & \int  {\cal D}\sigma {\cal D}\vec{\pi}
\exp \{- \int d^4 x {\cal L}_B + \nonumber\\
&& -\frac{1}{2} \int_{1/\Lambda^2}^{\infty} \frac{d\tau}{\tau}
\int d^4 x\; tr \langle x| e^{-\tau \left( -\partial_E^2 +g^2 MM^+ +
g \gamma\cdot(\partial M^+) \right)} | x \rangle \}\quad.\nonumber
\ea 
We remark that the lower boundary of the
integral over the proper time $\tau$ reflects the
ultraviolet cutoff scale $\Lambda$, 
while the upper limit of the proper
time integral represents the infrared scale.
Later, we will choose an
infrared cutoff function in such a way that no
ultraviolet cutoff in the proper time integration is needed.
Details concerning the heat kernel representation 
and definitions can be found in ref.~\cite{heat}.

Because we are interested in the calculation of the effective potential
no wave function renormalization corrections are taken into account.
This means that we set the wave function renormalization constant $Z_k
= 1$, which simplifies further calculations because  derivatives 
resulting from 
the heat kernel expression can be omitted. This is not a principal 
restriction of the method. On the contrary the chosen 
heat kernel regularization is designed
for a derivative expansion  and   
higher derivatives can be taken
into account not only for flat manifolds but also for curved
manifolds~\cite{ball}.

We find for the partition function at zero temperature the 
following expression which is exact up to the omission of derivatives 
$\partial_\mu \phi$ in the fermion exponent
\be
Z[J=0]  = 
\int  {\cal D}\sigma {\cal D}\vec{\pi}
\exp \{ -\frac{1}{2} \int_{1/\Lambda^2}^{\infty}
\frac{d\tau}{\tau}
\int d^4 x\, tr \langle x| e^{-\tau \left( -\partial_E^2 +g^2 \vec{\phi}^2 
\right)} | x \rangle -\int d^4 x  {\cal L}_B  \}\ . 
\ee
The fermion determinant is in general part of the functional
integration over the remaining meson fields weighted by the meson
Lagrangian ${\cal L}_B$. In the following we 
limit this integration over the fermions to the
one-loop level. Thus we decouple the effect of modified
fermions back on the meson dynamics, pull the non-local determinant in
front of the meson integration and replace the meson fields
$\vec{\phi}^2$ with the vacuum expectation value (VEV) 
$\phi_0^2$. The total effective action
splits into two pieces and becomes a sum of fermionic and bosonic
terms each depending on $\phi_0$:
\ba\label{gammatot}
-\ln Z =\Gamma({\phi_0})=\Gamma^F ({\phi_0})+\Gamma^B ({\phi_0}).
\ea
In a plane wave basis for the diagonal part of the heat kernel 
\cite{heat} the
effective action for the fermions $\Gamma^F$
has the form
\ba
\Gamma^F ({\phi_0} )& = & \frac{1}{2}\int d^4 x
\int_{1/\Lambda^2}^\infty 
\frac{d\tau}{\tau}
\int \frac{d^4 q}{(2\pi)^4} \left\{ tr_{{N_c}{N_f}\gamma}
e^{-\tau (q^2 + g^2 \phi_0^2)} \right\}\quad,
\ea
where the trace runs over color-, flavor- and spin-space and gives  a
factor $8N_c$ for two quark flavours.

The remaining meson integration is also performed on a one loop level and
is obtained by a saddle point approximation. The classical mesonic potential 
$V_0=  {m_0^2} \vec{\phi}^2 /2 + {\lambda_0}(\vec{\phi}^2)^2 /4 - c
\sigma$    
is expanded around the minimum, which corresponds to
the VEV $\phi_0$. Thus taking quadratic fluctuations 
of the mesons into account and
using again a heat kernel regularization for the fluctuation
determinant one finds 
the effective action for the bosons:
\ba\label{actionbose}
\Gamma^B ({\phi_0}) & = & -\frac{1}{2}\int d^4 x
\int_{1/\Lambda^2}^\infty 
\frac{d\tau}{\tau}
\int \frac{d^4 q}{(2\pi)^4} \left\{ tr_N 
e^{-\tau (q^2 + \frac{\partial^2 V_0}{\partial \phi_i \partial
    \phi_j})} \right\}\quad.
\ea
Here the trace runs over the  $(4 \times 4)$ fluctuation matrices in
$(\sigma,\vec{\pi})$-space.  
Note the opposite sign of the bosonic and fermionic
effective actions.
In the $\sigma$-model the calculation of the second 
derivatives of the potential $V_0$ 
is straightforward and the trace can be evaluated explicitly 
for $N (=4) $ components. One has to determine the
eigenvalues of the fluctuation matrix evaluated at the minimum of
the potential i.~e.~the masses of the system. With an
explicit symmetry breaking term $c$ the trace is given by
\ba\label{tracemaple}
\left. tr_N 
e^{-\tau \frac{\partial^2 V_0}{\partial \phi_i \partial \phi_j} }
\right|_{\vec{\phi}^2 = \phi_0^2}
& = & 3 e^{-\tau c / \phi_0}
+ e^{-\tau (2 \lambda_0 \phi_0^2 + c/ \phi_0)}\quad.
\ea
The first term on the r.h.s. of eq. (\ref{tracemaple}) represents the
mass squared eigenvalues associated with the three pions while the other term
describes the $\sigma$-meson (cf.~eqs.~(\ref{pionmass})-(\ref{quarkmass})). 
In general this trace is a sum over
all exponentiated squares of mass eigenvalues times the proper time $\tau$.

 
\subsection{The Renormalization Group Improvement}

The effective action $\Gamma (\phi_0)$, 
eqs.~(\ref{gammatot})-(\ref{actionbose}),
contains so far only fluctuations up 
to the one loop level 
and depends on the ultraviolet heat kernel cutoff $\Lambda$ 
via the proper time integration. 
In order to cut off the fluctuations in the infrared 
we introduce into the proper time integrand a universal $k$-dependent
function  
\ba
f_k ( \tau  ) &=& g(x)\\
x &=& \tau k^2
\ea
with a dimensionless
argument $x = \tau k^2$ 
which depends on 
the product of the proper time and the square of the 
infrared cutoff $k$. This
function $f_k$ has to 
satisfy three conditions: 

We require that
$g ( x \to 0 )  =f_k ( \tau k^2 \to 0  )=  1$, since the
action $\Gamma_k$ with infrared cutoff  should tend to the full
effective action $\Gamma$ at $k=0$, i.~e.~in
the limit $k \to 0$ the infrared cutoff is removed.

The function $g$  must satisfy
$ g ( x \to \infty ) \to 0$, to suppress modes with fixed ``virtuality'' $k$
for large $\tau$. Remember that the upper limit of the
proper time integration represents the infrared regime of the
effective action which is 
regularized by this condition of the function $g$.

Thirdly, the first derivative of the cutoff function
$g$ has to obey
\ba
g ' (x) & = & -x^2 h(x)
\ea
with $h(x)$ being any regular function in the vicinity of the origin.
This condition  assures that the renormalization group equations for the 
$k$-dependent effective action  are ultraviolet finite. 
The cutoff function regularizes the 
derivative of the effective potential with
respect to the infrared parameter $k$, therefore a 
further ultraviolet heat kernel cutoff $\Lambda$ is unwanted.
Due the above condition the ultraviolet divergences in the 
effective action itself are not affected by the presence of the infrared
cutoff \cite{wet1,flor}.
It is this condition which distinguishes the cutoff function from
a simple minded mass cutoff.

One possible choice for the function $g (x)$ which satisfy all these 
requirements is
\ba
\label{cutoff}
g (x)  =  e^{-x} ( 1 + x + \frac{1}{2} x^2 )
\ea
The function $g$
depicts a smooth cutoff of the momentum modes and allows for an
elegant and simple form
of the evolution equations. 

A smooth cutoff is unavoidable if one wants to have a well
defined derivative expansion. For the effective potential alone also a
sharp cutoff would do \cite{wegn},\cite{hase},\cite{morri}. 

Even in this case the above choice has
great advantages over a sharp cutoff, at least in 
the proper-time integral formalism where it may cause 
unpleasant oscillations (cf.~e.g.\cite{feld}).

The effective action with the cutoff function $f_k (\tau)$
has now the following form:
\ba
\label{potential}
\Gamma_k [{\phi_0}] & = & \int d^4 x V_k ({\phi_0})
\ea
with
\ba
\label{potential2}
V_k   =  - \frac{1}{2} \int\limits_{0}^\infty
\frac{d\tau}{\tau} f_k (\tau)
\int \frac{d^4 q}{(2\pi)^4} \left\{ tr_N e^{-\tau (q^2 +
\frac{\partial^2 V_0}{\partial \phi_i \partial \phi_j} )}
- tr e^{-\tau (q^2 + g^2 \tilde{\vec{\phi^2}} )} \right\}\quad.
\ea
This equation defines the  one loop effective 
potential with an additional infrared cutoff function.
In a further approximation
we truncate the effective potential at terms $\propto (\vec{\phi}^2)^2$.
Such a procedure leads to simple evolution differential equations for few 
couplings ($\lambda, m^2$). 
These evolution equations are 
improved by replacing the bare 
couplings ($\lambda_0 ,m_0^2$) and the constant VEV ($\phi_0$) with
the running couplings ($\lambda_k ,m_k^2$) and 
the running VEV ($\phi_k$). This renormalization group improvement 
resums the relevant infrared divergent Feynman diagrams and gives the 
important nonperturbative evolution of the potential. 
The
substitution can be compared with ordinary renormalization of $n$-point
functions in perturbation theory,
where one replaces on the one loop level
the bare couplings with the renormalized ones. This simple
strategy is pursued
in this paper and has many advantages. 
Another possibility is to derive 
an evolution equation for the full potential using the heat kernel 
regularization. 
This is postponed to 
another paper in the future \cite{papp}.

The evolution of the potential
itself with virtuality $k$ leads to two
phases and therefore two sets of equations.
In the chiral limit  the minimum of the potential (VEV)
will evolve from a large value to  zero 
in a second order phase transition.
For finite pion masses $( c\neq 0)$ this transition
becomes a crossover and the 
two sets of flow equations merge to one set of equations, as we will 
show analytically. 
In order to simplify and clarify the following 
discussion we start with the chiral limit and
set $c =0$ for the rest of this section.

For high virtuality $k$ the system is in the symmetric phase which is
defined by a vanishing VEV ($\phi_k =0$)
of the potential. Through the derivative with respect to
the fields
\ba
V_k ' & := & \frac{\partial V_k }{\partial {{\phi}^2}}
\ea
we can define the mass $m^2_k$ and the coupling constant $\lambda_k$
in the symmetric regime 
\ba
\frac{m^2_k}{2} & := & V_k ' ( \vec{\phi^2}=\phi^2_k =0 )\quad,\nonumber\\
\frac{\lambda_k}{2}  & := & V_k '' ( \vec{\phi^2}=\phi^2_k =0 )\quad.
\ea
The effective potential at the minimum is given by
\ba
v_k := V_k ( \vec{\phi^2}=\phi^2_k =0 )\quad.
\ea

Taking the derivatives of equation (\ref{potential2}) 
with respect to the scale $k$ we find
the following coupled sets of flow equations for the  
symmetric phase ($\phi^2_k = 0$):
\ba\label{eq21}
k \frac{\partial v_k}{\partial k} & = & k \frac{\partial V_k  (0)}
{\partial k}\quad,\\
\label{eq22}
\frac{k}{2} \frac{\partial m^2_k}{\partial k} & = & k \frac{\partial V_k ' (0)}
{\partial k}\quad,\\
\label{eq23} 
\frac{k}{2} \frac{\partial \lambda_k}{\partial k} & = & k
\frac{\partial V_k '' (0)} 
{\partial k}\quad.
\ea

In the spontaneously broken region the VEV is finite ($\phi_k \neq 0$)
and the mass parameter $m^2_k$ tends to negative values. 
We prefer to parameterize the evolution of the potential in this
region in terms of $\lambda_k$ and the minimum of the potential 
$\phi_k$, which is defined by
\ba
V_k ' ( \phi_k ) &=& 0\quad.
\ea
This equation enables us to find the evolution of the minimum $\phi_k$.
By again taking the derivatives with respect to the scale $k$ we find
the flow equation for $\phi^2_k$ in the
broken phase ($\phi^2_k \neq 0$)
\ba
k \frac{\partial v_k}{\partial k} & = & 
k \frac{\partial V_k  (\phi^2_k )}
{\partial k}\quad,\\
\frac{k}{2} \frac{\partial \phi^2_k}{\partial k} & = &
- \frac{k}{2 V_k '' (\phi^2_k )} \frac{\partial V_k ' (\phi^2_k )}{\partial k}
= - \frac{k}{\lambda_k} \frac{\partial V_k ' (\phi^2_k )}{\partial k}\quad,\\ 
\frac{k}{2} \frac{\partial \lambda_k}{\partial k} & = & 
k \frac{\partial V_k '' (\phi^2_k )}
{\partial k}\quad,\qquad \mbox{if}\quad V_k ''' = 0\quad.
\ea

Here an important remark must be added, 
that we first evaluate the derivatives on the right hand
sides for constant couplings $\lambda_0$, $m^2_0$ and constant 
VEV $\phi_0$ with respect to $k$.
Thus the scale derivative acts only on the infrared cutoff function 
$f_k (\tau)$ inside the proper time integrand.


\section{Equation of state in the chiral limit}
\label{secchirallimit}

In this section we discuss the flow equations for the 
two flavour linear sigma model in the chiral limit.
We present the numerical solution of the flow equations
with respect to the scale $k$ and generalize 
the equations to finite temperature.

\subsection{The evolution for $T=0$}

In the symmetric phase at zero temperature we use the
equations (\ref{eq21},\ref{eq22},\ref{eq23}) and get the evolution equations
for the free energy density, mass parameter and quartic coupling. Due to
the choice of the cutoff function $f_k(\tau)$ the resulting equations
are very transparent. The heat kernel cutoff is not a simple
mass cutoff, in effect the derivative cf.~eq.~(\ref{cutoff})
of the cutoff function
is simpler than the cutoff function itself, therefore
the evolution equations have such a concise form:  

\ba\label{1flowsym}
k \frac{\partial v_k}{\partial k} &= & - \frac{20}{ 2 (4\pi )^2} k^4\quad,\\
\label{2flowsym}
\frac{k}{2} \frac{\partial m^2_k}{\partial k} & = &
-\frac{3 \lambda_k k^2}{(4 \pi)^2} 
\frac{1}{\left( 1 +  m^2_k/k^2 \right)^2}
+ \frac{4 N_c g^2 k^2}{(4 \pi)^2}\quad,\\
\label{3flowsym}
\frac{k}{2} \frac{\partial \lambda_k}{\partial k} & = &
\frac{12 \lambda_k^2}{(4 \pi)^2} 
\frac{1}{\left( 1 + m^2_k/k^2 \right)^3}
- \frac{8 N_c g^4}{(4 \pi)^2}\qquad.
\ea

The factor $20$ in eq.~(\ref{1flowsym}) 
arises from the combination $4 - 8N_c$ (4 mesons minus $4 N_f N_c = 
8 N_c$ fermionic degrees of freedom). In the symmetric
phase the pions are degenerate with the $\sigma$-meson
and the squared mass parameter is always positive. The VEV
$\phi_k$ vanishes and does not appear on the right hand sides
of the evolution equations. 
In the broken phase the pion mass vanishes in the chiral limit.
The constituent quark mass is proportional to $g \phi_k$ and is zero in 
the symmetric phase.

\begin{figure}[hbt]
\unitlength1cm
\begin{center}
\begin{picture}(15,9)(-1,-0.5)
\put(2,7){Quartic coupling $\lambda_k$}
\put(5.5,-0.5){$k$ [MeV]}
\includegraphics{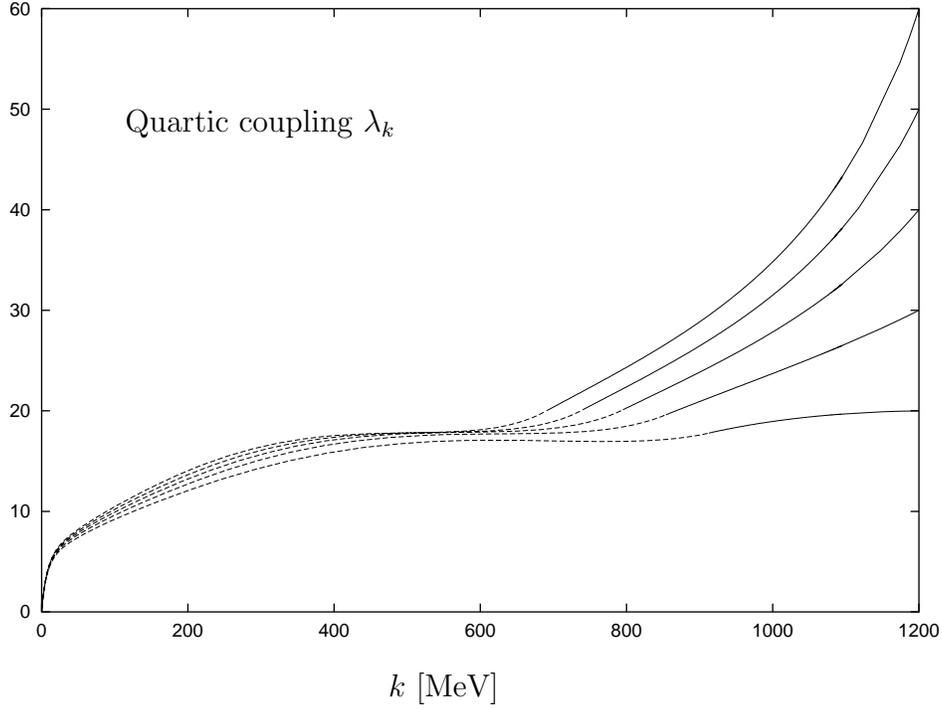}
\end{picture}
\parbox{12cm}
{\caption{\label{fig1} The evolution of $\lambda_k$ with 
respect to the scale $k$ for different initial values.}}
\end{center}
\end{figure}

To solve the evolution equations we need three initial values 
of $v_k$, $\lambda_k$ and $m_k^2$ at $k=\Lambda$ in the UV.
We start the evolution in the symmetric phase deep in the ultraviolet
region of QCD at $\Lambda = 1.2$ GeV and set $v_k = 0$.
The other two parameters $\lambda_k$ and $m_k^2$ are adjusted to get
$\phi_{k=0} = 93 \mbox{MeV} = f_\pi$ at $k=0$ and the transition to chiral
symmetry breaking at $k = k_{\chi SB} = 0.8$ GeV. This leads to 
$m^2_{k=\Lambda} = (550 \mbox{MeV} )^2$ and
$\lambda_{k=\Lambda} = 40$.

In addition we need the Yukawa coupling $g$ 
which we fix to $g = 3.2$ for all $k$ in order to
get a constituent quark mass of $300$ MeV
at the end of the evolution. 
Because we do not take any wave function renormalization into account 
$g$ has only
a weak scale dependence.
In the work of Wetterich et al.~\cite{wett} the main contribution
to the strong evolution of the Yukawa coupling stems from the 
evolution of the meson wave function renormalization.

In fact, the initial value for the quartic coupling 
$\lambda_{k=\Lambda}$ 
turns out to be not very important due to the Gaussian fix point
$\lambda( k \to 0 ) \to 0$.
This behaviour can be seen in figure \ref{fig1}, where the evolution of 
the quartic coupling towards zero for different initial values
is plotted.
Near $k=\Lambda$ the evolution of $\lambda_k$ is dominated 
by the bosonic part, the first term on the r.h.s. of eq.~(\ref{3flowsym})
whereas the evolution of $m^2_k$ responds to the fermion loops.
When we reach chiral symmetry breaking at the scale 
$k = k_{\chi SB} \approx 0.8$ GeV, we switch to the 
equations for the broken phase characterized by $\phi^2_k \neq 0$:

\ba\label{1flowbroken}
k \frac{\partial v_k}{\partial k} &=& \frac{k^4}{2 (4\pi )^2}
\left[ 3 + \frac{1}{(1+ 2\lambda_k \phi_k^2 / k^2)}
- \frac{8N_c}{(1+g^2 \phi^2_k / k^2)}\right]\ ,\\
\label{2flowbroken}
\frac{k}{2} \frac{\partial \phi^2_k}{\partial k} & = &
\frac{3 k^2}{2 (4 \pi)^2} \left[ 1 + 
\frac{1}{\left(1+2\lambda_k \phi^2_k / k^2\right)^2} \right]
- \frac{4 N_c}{(4\pi)^2}
\frac{k^2g^2}{\lambda_k}
\left[\frac{1}{\left(1+g^2 \phi^2_k / k^2\right)^2}\right]\ ,\\
\label{3flowbroken}
\frac{k}{2} \frac{\partial \lambda_k}{\partial k} & = &
\frac{3 \lambda_k^2}{(4 \pi)^2} \left[ 1 +
\frac{3}{\left(1+2\lambda_k \phi^2_k / k^2\right)^3} \right]
- \frac{8 N_c}{(4\pi)^2} g^4
\left[
\frac{1}{\left(1+g^2 \phi^2_k / k^2\right)^3}\right]\quad.
\ea

The first two terms on the r.h.s. of the above equations are 
again related to the boson fluctuations while the last parts are connected 
to the fermionic contributions to the evolution. 

\begin{figure}[hbt]
\unitlength1cm
\begin{center}
\begin{picture}(15,9)(-1,-0.5)
\put(5.5,-0.5){$k$ [MeV]}
\put(3,2.5){$\phi_k$}
\put(9,6){$m_k$}
\put(-0.5,8){[MeV]}
\includegraphics{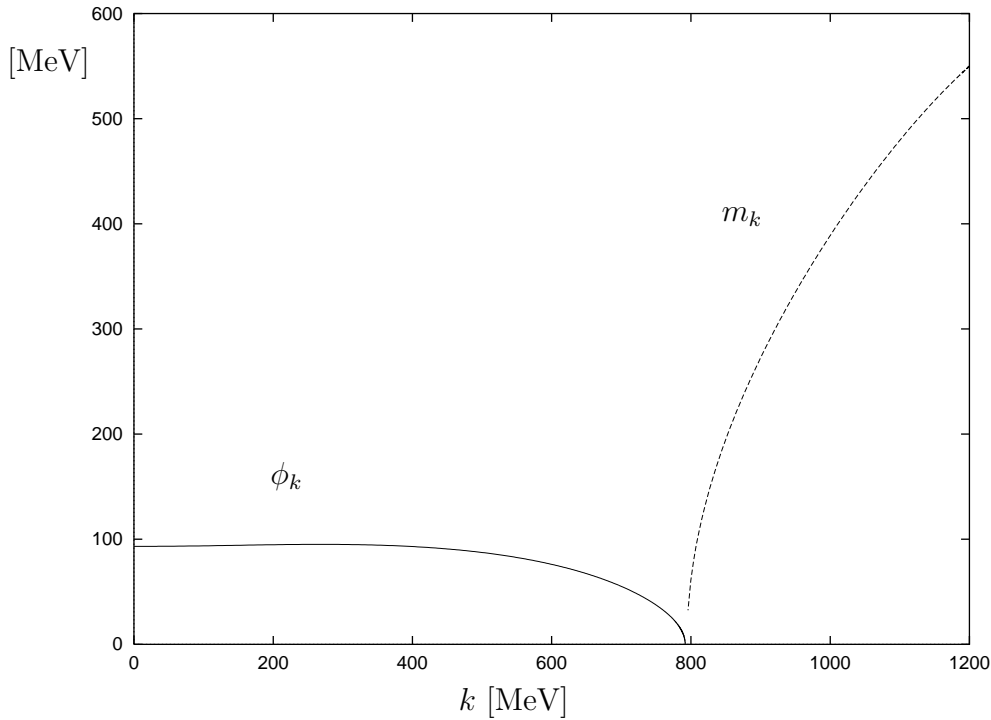}
\end{picture}
\parbox{12cm}
{\caption{\label{fig2} The evolution of the minimum 
of the effective potential $\phi_k$ for $k<k_{\chi SB}$ 
and the evolution of the mass parameter $m_k$ for $k>k_{\chi SB}$.}}
\end{center}
\end{figure}

The constant factor in the first part of the upper equations 
comes from 
the three massless pions while
the second part
of the bosonic term contains the contribution
of the $\sigma$-meson with squared mass $2\lambda_k \phi^2_k$.
In the broken phase the $\sigma$-meson is massive ($\phi_k \neq 0$) 
and the $\sigma$-meson contribution decouples from the evolution
for small $k$ values in the IR region, when $k \ll 2\lambda_k
\phi_k^2$; whereas the massless pions 
contribute permanently
to the flow in the chiral limit.
The fermionic quark part of the flow equation 
enters with the opposite sign to the evolution and the quarks 
become more and more 
massive in the infrared. The factors have a similar origin as in the
symmetric phase. The contribution from the pions is three
times larger 
than the $\sigma$-meson  contribution. Two flavors , two particles and 
anti-particles and two spin degrees of freedom yield a factor of
$8N_c$.

The functions
in squared brackets can be called threshold functions, because
they describe the smooth decoupling of massive modes from the
evolution towards the infrared limit of the theory given by
the scale $k$. 
With the heat kernel method and our choice of the function 
$f_k (\tau)$ we can calculate these threshold functions
analytically. They are of the form
$(1+ \mbox{mass}^2/k^2)^{-p}$ 
where $p$ is some positive integer power.
So the quark loop gives a factor
$(1 + g^2\phi_k^2/k^2)^{-p}$ with positive $p$ which switches off the
quark contribution to the evolution of the couplings for $k \ll g\phi_k$.
Similarly, the square of the sigma mass over the
infrared scale $2 \lambda_k \phi_k^2/k^2$
suppresses the massive boson loop. At small evolution parameters
only zero mass pion loops drive the evolution. This 
is very intuitive and coincides with the ideas of chiral perturbation theory.
In fact, it has been also shown \cite{dir2} that 
this model reproduces the
chiral expansions even beyond leading order. 
As we will see in the equation of state the suppression of the
quark terms for  the evolution does not mean that the quarks do
not contribute to the thermodynamic functions like the 
pressure or energy density below the phase transition temperature.
In this respect the model has the same defects as more 
naive NJL models without quark confinement.

We remark that in the heat kernel expression for the
effective potential the inverse fermion propagator enters
quadratically in the combination $D D^+$ therefore we can use the same
infrared cutoff function $f_k$ as in the bosonic integral 
without breaking chiral symmetry. If we want to 
take into account the running of the Yukawa coupling $g$ 
the fermion propagators also have to be squared.
One also recognizes the signs of the bosonic and fermionic
contributions to the $\beta$-function. The bosons lead to an infrared
stable (ultraviolet unstable) $\lambda$-coupling, 
whereas the fermions counteract this
tendency. Going from high $k$ to low $k$ one sees that the 
mesonic self-interaction $\lambda_k$ will balance at intermediate values
of $k$, whereas in the far infrared the boson term wins 
(cf.\ figure~\ref{fig1}). The chiral symmetry
changes at $k=k_{\chi SB} \approx 800$ MeV, where the
system goes over from the symmetric phase to the spontaneously broken
phase. At this scale $k$ 
eqs.~(\ref{3flowsym}) and (\ref{3flowbroken}) become identical, 
i.e. the $\beta$-function of
$\lambda_k$ is continuous (cf. 
figure~\ref{fig1}).

We have argued in a separate paper \cite{pirn} that this
transition at zero temperature may be visible in electron scattering,
where real photoproduction  with virtuality $Q^2=0$
corresponds to a heat kernel cutoff $k^2=0$ and inelastic scattering
with higher $Q^2$ to the equivalent $k^2$ scales.
The transition with increasing $k$ from the constituent quark to the
parton in deep inelastic scattering is a natural consequence in the
linear sigma model.
The vacuum expectation value $\phi_k$ stabilizes at small
values of $k$ and the evolution ends with 
$\lim_{k \to 0} \phi_k = f_\pi$.
When the heavy particles, the sigma and quarks, have
decoupled, the change of the vacuum expectation value becomes proportional to
$k$, the evolution stops (see figure~\ref{fig2}).
It is interesting to investigate 
how a more sophisticated calculation with running
Yukawa coupling $g_k$ and wave function renormalization $Z_k$
relates  the field theoretic model to more observables.
Thereby the evolution itself may be tested in a region which
perturbative QCD i.~e.~the Altarelli Parisi evolution equations cannot reach.


\subsection{The evolution for finite $T$}

The extension of the flow equations to finite temperature 
for thermal equilibrium is achieved by the Matsubara technique.   
We integrate the momenta in equation (\ref{potential2}),
by splitting the zero momentum component from the three-dimensional 
spatial momentum components
and convert the integration over $q_0$ into a summation over
Matsubara frequencies $\omega_n$ for the bosons and $\nu_n$
for the quarks \cite{kapu}:
\ba
\omega^2_n &=& 4\pi^2 n^2 T^2\quad,\\
\nu_n^2 &=& (2n+1)^2\pi^2T^2\quad;\qquad n \in Z\!\!\!Z\ .
\ea 

In the symmetric phase the corresponding equations are written 
in terms of the positive mass parameter $m^2_k$ and $\lambda_k$:
\ba\label{1flowtsym}
k \frac{\partial v_k}{\partial k} &=& \frac{k^4}{2 (4\pi )^2}
\frac{\pi T}{k}\left[ 4\sum_{n=-\infty}^\infty
\frac{1}{(1+\omega_n^2/k^2 )^{3/2}}
-8N_c \sum_{n=-\infty}^\infty
\frac{1}{(1+\nu_n^2/k^2 )^{3/2}}\right] ,\\
\label{2flowtsym}
\frac{k}{2} \frac{\partial m_k^2}{\partial k} & = &
-\frac{3 \lambda_k^T}{(4 \pi)^2} k^2\left[
\frac{3\pi}{2} \frac{T}{k}
\sum_{n = -\infty}^\infty
\frac{1}{\left( 1 + \left(\omega^2_n+m^2_k\right)/{k^2} 
\right)^{5/2}}\right]\nonumber\\
&&\nonumber\\
&& + \frac{4 N_c g^2}{(4\pi)^2} k^2\left[\frac{3\pi}{2}\frac{T}{k}
\sum_{n = -\infty}^\infty
\frac{1}{\left( 1 + \nu^2_n/k^2 \right)^{5/2}}\right]\quad,\\
&&\nonumber\\ 
\label{3flowtsym}
\frac{k}{2} \frac{\partial \lambda_k^T}{\partial k} & = &
\frac{12 (\lambda_k^T)^2}{(4 \pi)^2}
\left[\frac{15\pi}{8}\frac{T}{k}
\sum_{n = -\infty}^\infty
\frac{1}{\left( 1 + \left(\omega^2_n+m^2_k\right)/{k^2} 
\right)^{7/2}}\right]\nonumber\\
&&\nonumber\\
&& -\frac{8 N_c g^4}{(4\pi)^2}
\left[\frac{15\pi}{8}\frac{T}{k}
\sum_{n = -\infty}^\infty
\frac{1}{\left( 1 + \nu^2_n/k^2 \right)^{7/2}}\right]\quad.
\ea

The flow equation for the free energy density $v_k (0)$
in the symmetric phase, eq.~(\ref{1flowtsym}), does not depend on any
coupling neither at zero nor at finite temperature. 
The minimum of the effective potential  yields an integration constant.
Each contribution to the flow equations at finite temperature
is taken into account in a similar way as in the zero temperature
case (see above). 
The modifications in the flow equations appear 
in the threshold functions within the squared brackets in 
eqs.~(\ref{1flowtsym}) - (\ref{3flowtsym}). 

\begin{figure}[hbt]
\unitlength1cm
\begin{center}
\begin{picture}(15,9)(-1,-0.5)
\put(5.5,-0.5){$T/k$}
\put(6,5){$y=0$}
\put(8,3.5){$y=2$}
\put(9,2.5){$y=4$}
\put(2,8){$\frac{15 \pi}{8}\frac{T}{k}
\sum\limits_n
\frac{(1+y^2)^3}{\left( 1 + \omega^2_n/k^2 + y^2\right)^{7/2}}$}
\includegraphics{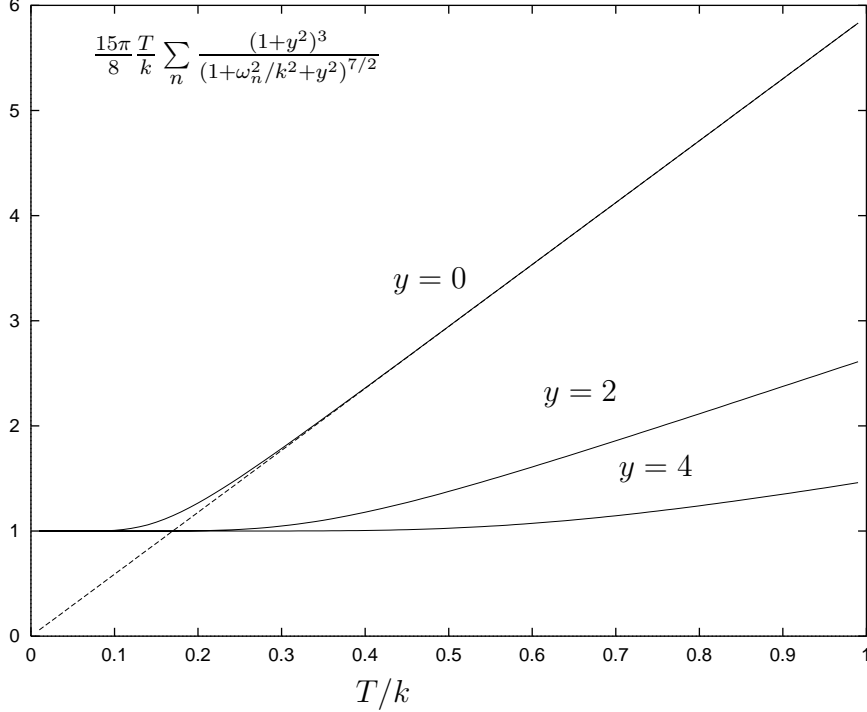}
\end{picture}
\parbox{12cm}
{\caption{\label{fig3} The ratio of bosonic threshold functions for
temperature $T=0$ and $T \neq 0$ with 
different mass parameters $y=0,2,4$ as function of $T/k$. 
The dashed line is the function $15 \pi /8\cdot T/k$ which 
demonstrates the linear behaviour of the threshold function for
large $T/k$ in the case of $y=0$.}}
\end{center}
\end{figure}

For the broken phase we 
get the following expressions:
\ba\label{1flowtbroken}
k \frac{\partial v_k}{\partial k} &=& \frac{k^4}{2 (4\pi )^2}
\frac{\pi T}{k}\left[ 3\sum_{n=-\infty}^\infty
\frac{1}{(1+\omega_n^2/k^2 )^{3/2}}
+ \sum_{n=-\infty}^\infty
\frac{1}{(1+(\omega_n^2 + 2 \lambda^T_k (\phi^T_k)^2)/k^2 )^{3/2}}
\right.\nonumber\\
&&\left.
-8N_c \sum_{n=-\infty}^\infty
\frac{1}{(1+(\nu_n^2 + g^2 (\phi^T_k)^2)/k^2 )^{3/2}}\right]\ .\\
\label{2flowtbroken}
\frac{k}{2} \frac{\partial (\phi^T_k)^2}{\partial k} & = &
\frac{3 k^2}{2(4 \pi)^2}\left[\frac{3\pi}{2}\frac{T}{k}
\sum_{n = -\infty}^\infty\left\{
\frac{1}{\left( 1 + \omega^2_n/k^2 \right)^{5/2}} +
\frac{1}
{\left( 1 + 
\left(\omega^2_n + 2\lambda^T_k (\phi^T_k)^2\right)/{k^2}\right)^{5/2}}
\right\}
\right]\nonumber\\
&&\nonumber\\
&& -\frac{4 N_c}{(4\pi)^2} \frac{g^2 k^2}{\lambda^T_k}
\left[\frac{3\pi}{2}\frac{T}{k}
\sum_{n = -\infty}^\infty
\frac{1}{\left( 1 + 
\left(\nu^2_n+ g^2 (\phi^T_k)^2\right)/{k^2} \right)^{5/2}}
\right]\quad,\\
&&\nonumber\\
\label{3flowtbroken}
\frac{k}{2} \frac{\partial \lambda_k^T}{\partial k} & = &
\frac{3 (\lambda_k^T)^2}{(4\pi)^2}\left[
\frac{15 \pi}{8}\frac{T}{k}
\sum_{n = -\infty}^\infty
\left\{
\frac{1}{\left( 1 + \omega^2_n/k^2 \right)^{7/2}} +
\frac{3}
{\left( 1 + 
\left(\omega^2_n + 2\lambda^T_k (\phi^T_k)^2\right)/{k^2}\right)^{7/2}} 
\right\}
\right]\nonumber\\
&&\nonumber\\
&& -\frac{8 N_c}{(4\pi)^2} g^4 \left[
\frac{15 \pi}{8}\frac{T}{k}
\sum_{n = -\infty}^\infty
\frac{1}{\left( 1 + 
\left(\nu^2_n + g^2 (\phi^T_k)^2\right)/{k^2} \right)^{7/2}}
\right]\quad.
\ea
In  the limit $\phi_k \to 0$, i.e. at the chiral symmetry breaking scale 
$k_{\chi SB}$
the equations for
the symmetric and the broken phase are continuous.

\begin{figure}[hbt]
\unitlength1cm
\begin{center}
\begin{picture}(15,9)(-1,-0.5)
\put(5.5,-0.5){$T/k$}
\put(5,3){$y=0$}
\put(7,5.5){$y=2$}
\put(9,7){$y=4$}
\put(4,8){$\frac{15 \pi}{8}\frac{T}{k}
\sum\limits_{n}
\frac{(1+y^2)^3}{\left( 1 + \nu^2_n/k^2 + y^2 \right)^{7/2}}$}
\includegraphics{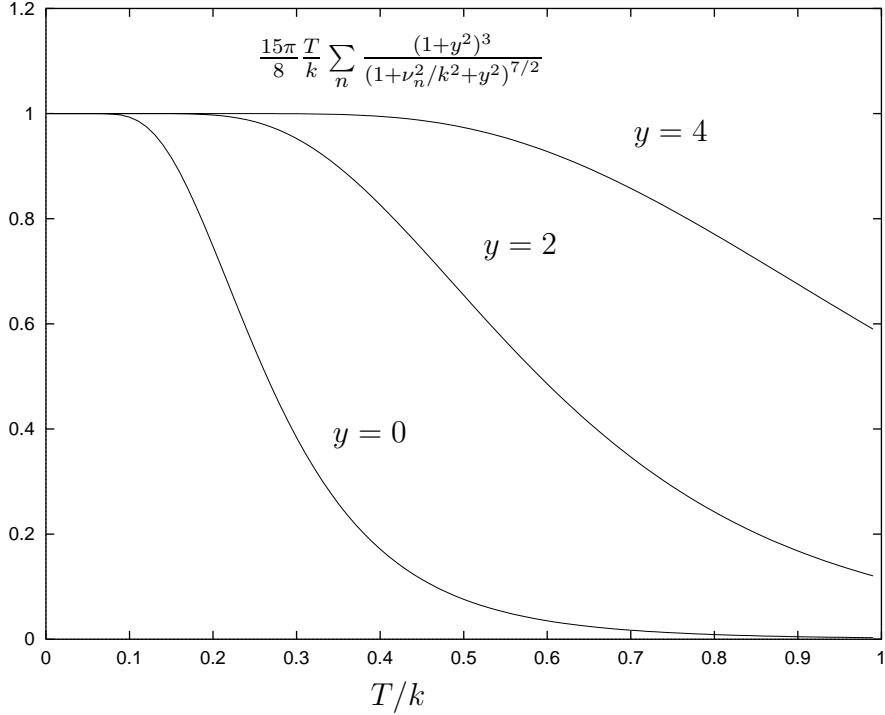}
\end{picture}
\parbox{12cm}
{\caption{\label{fig4} The ratio of fermionic threshold functions for 
$T=0$ and $T \neq 0$ with different
mass parameters $y=0,2,4$ as function of $T/k$.}}
\end{center}
\end{figure}

One sees that the equations arise from the same one loop
diagrams as the $T=0$ equations and therefore 
depend on the coupling constants in a similar way.
Due to  the three dimensional
momentum integrations fractional powers arise in the threshold functions.

In the limit of low temperatures $T/k \to 0$ we 
regain our old expressions at $T=0$, i.e. we have the relations: 
\ba\label{thresholdlimit}
\frac{3\pi}{2} \frac{T}{k}
\sum_{n = -\infty}^\infty
\frac{1}{\left( 1 + \left\{\omega^2_n \atop \nu^2_n \right\} 
/{k^2} +y^2\right)^{5/2}} 
& \to & \frac{1}{\left( 1 + y^2 \right)^2}
\ea

It is possible to analytically 
relate all Matsubara sums to the 
corresponding four dimensional integrals in the 
low temperature limit cf.~appendix~\ref{appmat} and ref.~\cite{stri}.
The above equations guarantee the right matching 
of the finite temperature equations
to the zero temperature equations, i.e.\ in the limit
$T \to 0$
the sets of equations for finite and zero temperature become identical.

We plot in figure~\ref{fig3} 
the ratios of finite temperature bosonic threshold function appearing
in the $\beta$-function for $\lambda_k^T$ over
the zero temperature $\beta$-function for $\lambda_k$.
Figure~\ref{fig4} shows the equivalent ratios of 
threshold functions for the fermions.
For small ratios $T/k$ up to $0.1$ the
threshold function ratio is constant which means that
for temperatures small compared to the infrared cutoff scale 
all Matsubara modes are important. The summation is effectively a
continuous integration.

For large ratios $T/k$ the 
bosonic threshold functions
increase linearly in $T/k$. (cf.\ figure~\ref{fig3}).
This increase is due to the $n=0$ Matsubara mode in the frequency sum.
In the large temperature limit the ($3+1$)-dimensional system reduces
to a $3$-dimensional system. Thus the phenomenon of
dimensional reduction   
is automatically included in the bosonic threshold functions.
In the fermionic case there is no zero  Matsubara frequency,
therefore
the fermionic threshold functions decrease with $T/k$ and
become zero. This means that for high $T/k$ the fermions decouple from
the further evolution.  
There also exists a stable plateau in the vicinity of the
origin for the Fermi-case as in the ref.~\cite{wett}.

The dimensional reduction manifests itself 
in the critical behaviour, as will be shown in the next section.


\subsection{Critical behaviour and exponents}
\label{seccritical}

The idea underlying the finite temperature calculation is the 
following: At large ultraviolet scale $k = \Lambda$ and finite temperature
$T$ the couplings of the effective theory are identical
to the $T=0$ case, since
the finite temperature only modifies the
boundary in the imaginary time direction but not the dynamics
on a small scale in space time.
This holds as long as the ratio $T/k < 0.12 $, 
i.e. in the region where
the ratio of the threshold functions of the ultraviolet
theory does not deviate strongly from unity 
(cf.~figures~\ref{fig3},\ref{fig4}).
Thus temperatures smaller than roughly $140$ MeV will not change the
$T=0$ couplings at finite temperatures at the ultraviolet scale
$\Lambda = 1.2$ GeV.

\begin{figure}[hbt]
\unitlength1cm
\begin{center}
\begin{picture}(15,9)(-1,-0.5)
\put(6,2.5){$\phi_k (T)$}
\put(9,6){$m_k (T)$}
\put(-0.5,8){[MeV]}
\put(2,2){$T = 5$ MeV}
\put(4.2,0.8){$T=160$ MeV}
\put(5.5,-0.5){$k$ [MeV]}
\includegraphics{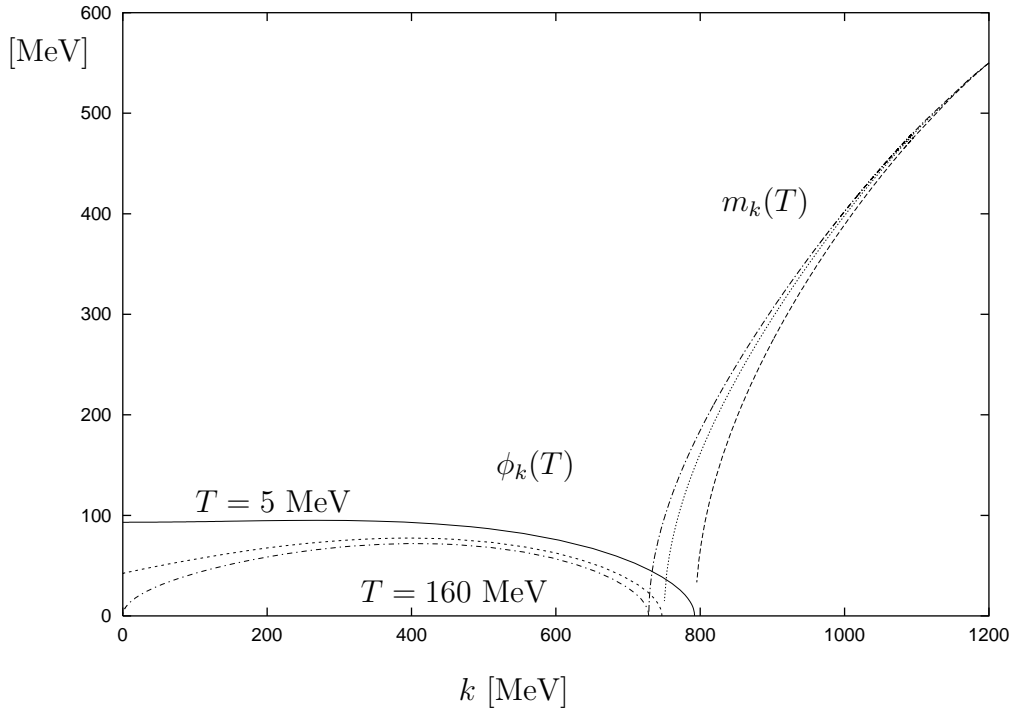}
\end{picture}
\parbox{12cm}
{\caption{\label{fig5} The VEV $\phi_k$ for $k<k_{\chi SB} (T)$ and the mass 
$m_k$ for $k>k_{\chi SB} (T) $ 
for different temperatures. (Upper line $T=5$ MeV, middle
line $T=150$ MeV and bottom line $T=160$ MeV). The decrease of the
order parameter $\phi_k$  
towards zero for $k=0$  signals the chiral phase transition.}}
\end{center}
\end{figure}

Therefore we  can use the same starting parameters as in  the $T=0$ theory to
obtain finite temperature results and solve
for each fixed temperature $T$ the evolution equations
as functions of $k$.  The numerical results show the following 
behaviour:

With increasing temperature the mass parameter $m_k$
decreases slightly 
more slowly towards the condensation point (cf.\ figure~\ref{fig5}),
but the main effect of the finite temperature occurs below 
$k\simeq 800$ MeV.

The threshold functions contain an additional damping due to the
Matsubara frequency. Below the condensation point the
boson condensate $\phi_k^2$ reaches less high  values at finite 
temperatures than 
at $T=0$. It decreases with $k \to 0$ for all temperatures.
This effect is due to 
the pion fluctuations. 
The critical temperature is reached when the condensate tends to zero.
The exact value of the critical temperature depends on the interplay
between the relevant sigma mass and quark mass in the intermediate
$k-$region. If the ``squared sigma mass''  $2\lambda_k \phi^2_k$ 
is small relative to $k^2$ 
the sigma fluctuations drive 
symmetry restoration together with the
pionic fluctuations. Our
values of the relevant parameter $2\lambda_k \phi^2_k$ 
at $T=0$ are typically around four times the quark
constituent mass squared evaluated between $100$ MeV $< k < 300$ MeV. 
The sigma mass itself is well defined for all $k$ in the case
of explicit symmetry breaking.
The particle data book \cite{pdgr} gives for the $f_0$-
or $\sigma$-meson a mass range between $400$ MeV and $1200$ MeV. From the
nucleon nucleon interaction, however,  we know that the 
intermediate  range attraction comes from an equivalent sigma mass 
which has a value of $\approx 560$ MeV 
e.g.~in the Reid soft core potential 
\cite{reid}.
We add as a note of caution that
even if
the $O(4)$ dynamics determine critical indices at $T_c$
correctly,
the value of  $T_c$ may be influenced by gluonic
degrees of freedom exterior 
to the linear $\sigma$-model in its present form. E.~g.~one can 
imagine that the
$\sigma$ couples to the glueball which dissolves at the critical 
temperature if the deconfinement transition occurs together
with the chiral phase transition.
With the linear sigma model as given and the above  parameters we obtain 
a critical temperature:
\ba
T_c \approx 164 \ \mbox{MeV}\ .
\ea 

This temperature is higher than the critical temperature 
of the Wetterich group, even 
if $f_\pi= 93$ MeV had been used in \cite{wett} for the chiral
limit. Wetterich et al.~obtained a critical temperature 
$T_c \approx 100$ MeV.
They also use different 
non-analytic threshold functions and
wave function and coupling constant renormalization. But we 
do not expect 
the latter modifications to
change the critical temperature drastically.

\begin{figure}[hbt]
\unitlength1cm
\begin{center}
\begin{picture}(15,9)(-1,-0.5)
\put(-0.8,8){$\phi_{k=0} (T)$}
\put(-0.7,7.2){[MeV]}
\put(5.5,-0.5){Temperature [MeV]}
\put(8.5,7){chiral pert.\ theory}
\put(10.5,6.5){\vector(0,-1){1.5}}
\put(5.9,3.9){$\times$ : $\left(\frac{\D T_c -T}{\D T_c}\right)^\beta$}
\includegraphics{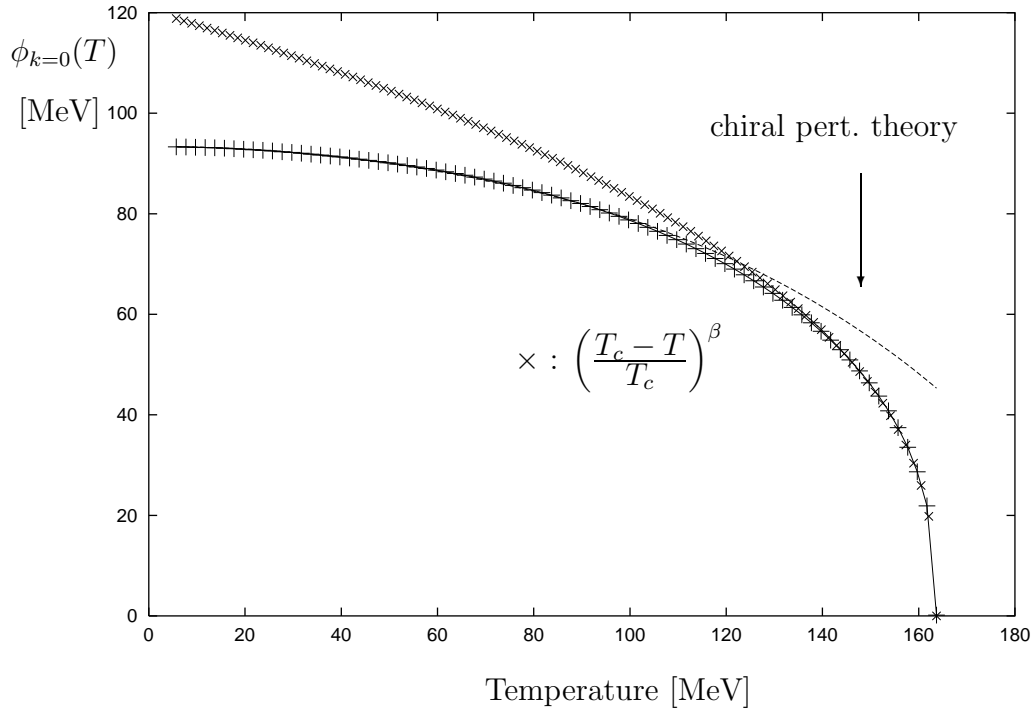}
\end{picture}
\parbox{12cm}
{\caption{\label{fig6} The order parameter $\phi_{k=0}$ as a function of the  
temperature $T$.}}
\end{center}
\end{figure}

In figure~\ref{fig6} we show the order parameter 
$\phi_{k=0} (T)$ 
as a function of temperature. At a first glance 
one might think that  the calculated behaviour
follows roughly  the 
chiral perturbation theory until $T \approx 110$ MeV, then it
deviates because of  strong massless mesonic fluctuations
which are not properly taken into account in chiral perturbation theory.
In detail our calculation coincides with chiral perturbation 
theory up to temperatures about $45$ MeV then small deviations are visible.
In chiral perturbation theory the temperature dependence of the 
light quark condensate for massless
quarks, plotted in figure~\ref{fig6}, is given by the
expression
\ba
\frac{\langle \bar{q} q\rangle_T}{\langle \bar{q} q\rangle_0}
& = & 1-\frac{T^2}{8 f_\pi^2} -\frac{T^4}{384 f_\pi^4}-
\frac{T^6}{288 f_\pi^6}\ln \frac{\Lambda_q}{T} + {\cal O}(T^8)
\ea
with $\Lambda_q = 470 \pm 110$ MeV \cite{chpt}.
In a model without quark confinement like the one used here  the fermion
loop may be overestimated cf.~\cite{wachs}. In general the mesonic modes
have Boltzmann suppression factors 
$\propto \exp ( - 2 m_Q/T)$ compared to a Boltzmann factor
$\propto \exp ( - m_Q/T)$ 
for the quark degrees of freedom, where $m_Q$ is the constituent quark
mass. Therefore the mesons probably contribute to the pressure at somewhat
higher temperatures 
than unconfined quarks.

The renormalization group 
calculation gives a non mean field behaviour for the
order parameter at $T_c$, since the long range fluctuations of the
softening $\sigma$ modes are gradually taken into account by 
the successive lowering of the cutoff.
The renormalization group flow equations 
at finite temperature sum the diverging
thermal fluctuations
properly near the critical point. 
 
\begin{figure}[hbt!]
\unitlength1cm
\begin{center}
\begin{picture}(15,9)(-1,-0.5)
\put(-1.,8){$log \ \frac{\D \phi_k}{\D \mbox{[MeV]}} $}
\put(5.5,-0.5){$log \ \frac{\D T_c - T}{\D T_c} $}
\put(2,8){Critical Exponent $\beta$}
\put(7.8,3.0){mean field}
\put(8.0,3.4){\vector(-1,1){1}}
\put(4.5,6.4){$\beta = 0.40$}
\put(5.0,6.2){\vector(1,-1){1}}
\includegraphics{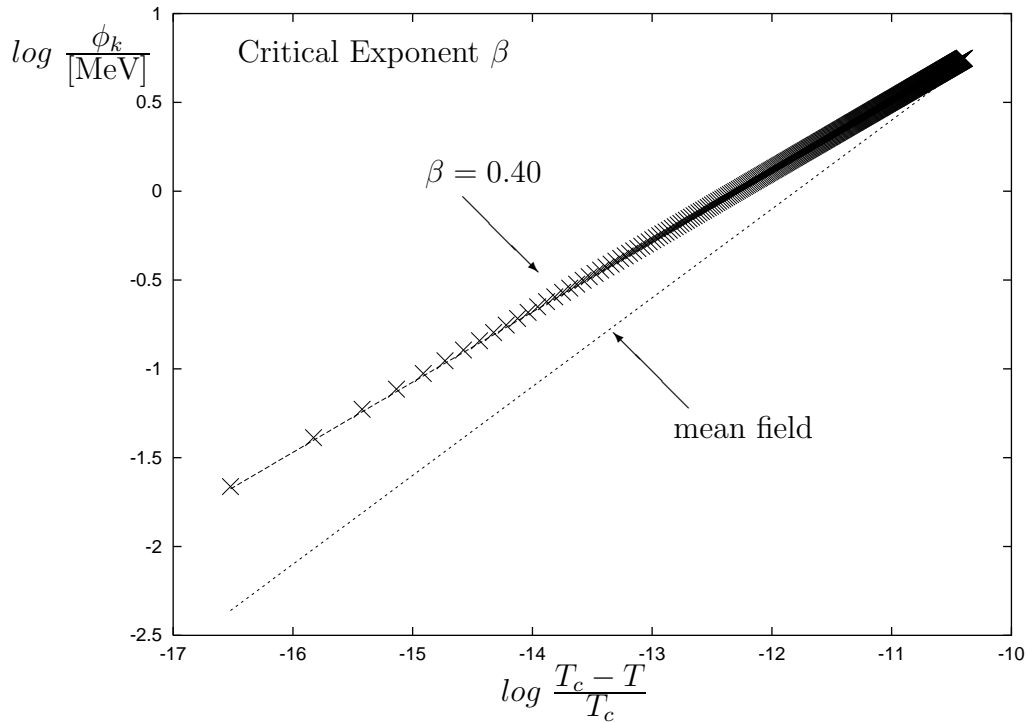}
\end{picture}
\parbox{12cm}
{\caption{\label{fig7} Determination of the critical exponent $\beta$
from the temperature dependence of the order
parameter $\phi_k$. }}
\end{center}
\end{figure}

In the vicinity of $T_c$ we  obtain  
scaling behaviour of the order parameter with a 
critical exponent $\beta \approx 0.40$ as shown in figure~\ref{fig6}.
In fact  it is possible to determine the exponent  $\beta$ 
by analyzing numerically 
the power
law behaviour of the order parameter near $T_c$.
We plot $log (\phi_k)$ versus 
$log ((T_c - T)/T_c)$ in figure~\ref{fig7}. 
The data points lie on a linear curve, thus $\phi_k$ scales like
\ba
\phi_k \propto \left| \frac{T - T_c}{T_c} \right| ^{\beta}
\ea
with a critical exponent $\beta \approx 0.40$ which is 
in good agreement 
with lattice results for the $O(4)$-theory
in three dimensions \cite{kana}. 
We also compare the result with the mean field value $\beta = 0.5$ 
in figure~\ref{fig7}.


\begin{figure}[hbt!]
\unitlength1cm
\centering \leavevmode
\begin{center}
\begin{picture}(15,9)(-1,-0.5)
\put(-2.5,8){$log \{ c(T_c) - c(T) \}$}
\put(5.5,-0.5){$log \ \frac{\D T_c - T}{\D T_c} $}
\put(3,8){Critical Exponent $\alpha$}
\put(7.2,3.5){$\alpha = - 0.395$}
\includegraphics{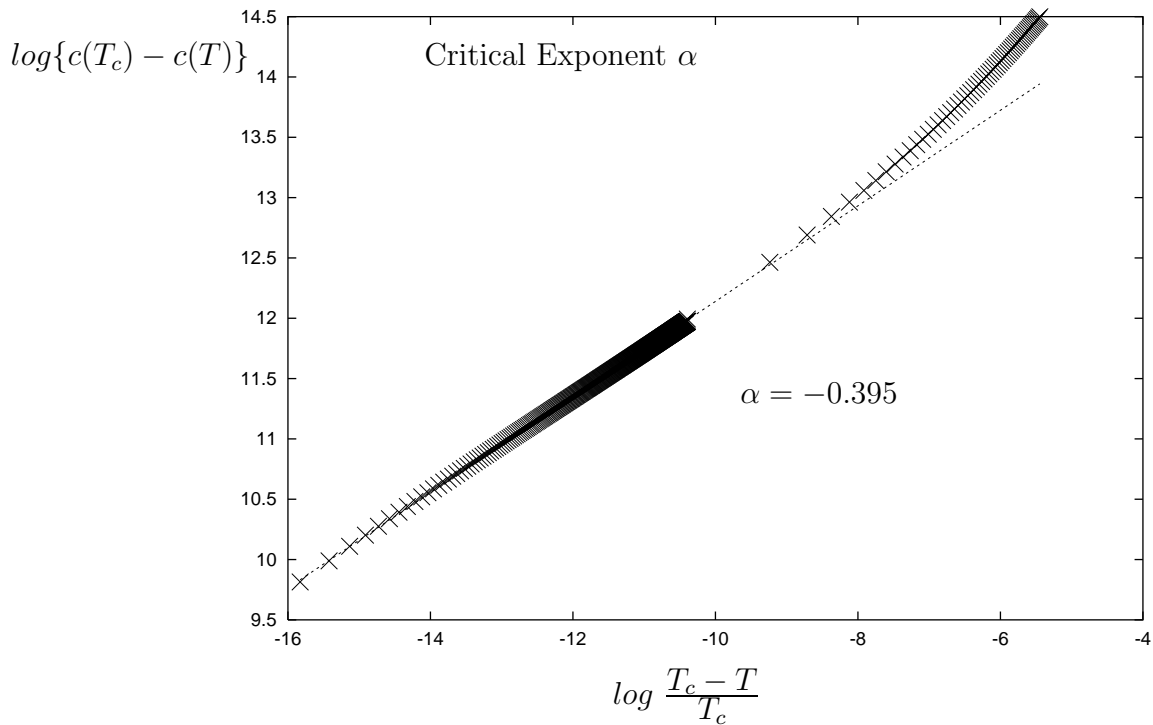}
\end{picture}
\parbox{12cm}
{\caption{\label{fig8} Determination of the critical exponent $\alpha$
for the specific heat $c(T)=c(T_c)+ const.|T-T_c|^{-\alpha}$ 
in the vicinity of the critical temperature.}}
\end{center}
\end{figure} 

At finite temperature the limit $T/k \to \infty$ 
enforces  dimensional reduction. One knows, that 
the linear sigma model lies  in the same universality class
as the three-dimensional $O(4)$-Heisenberg model, 
since the fermions are not contributing 
to the critical fluctuations near $T_c$ \cite{pis}.
Thus we deal with an effectively three-dimensional
bosonic $O(4)$-symmetric sigma model.
For the $O(4)$-theory there are six common critical exponents
but only two of them are independent because they are related by four 
well-known scaling relations of the three-dimensional scalar 
$O(4)$-model:
\ba
\alpha & = & 2 - d \nu \nonumber\ ,\\
\beta & = & \frac{\nu}{2} (d -2 + \eta )\nonumber\ ,\\
\gamma & = & (2 - \eta ) \nu \nonumber\ ,\\
\delta & = & \frac{d+2-\eta}{d-2+\eta}\nonumber\quad,
\ea
where $d$ is the dimension of the system ($d = 3$) \cite{binn}. 
Using these relations
our results for 
$\beta = 0.40$ and $\eta = 0$ (no wave function renormalization is used)
predict the values in tab.~(\ref{tab1}) which are denoted by RG.

\begin{table}[ht!]
\begin{center}
\begin{tabular}{c||c|c|c|c}
     & \quad  Mean Field \quad  	& \quad    RG  \quad   &  \quad  $O(4)$ Lattice \quad & \quad QCD Lattice 
\\[0.5ex]\hline \hline\rule[-0.5ex]{0mm}{6mm}
$1/\beta \delta $	&   2/3 	&  \quad 0.5 $\pm$0.06 \quad     & 0.537(7)& 0.77(14)\\
$1/\delta $    		&   1/3		&   0.2       & 0.2061(9)& 0.21 \ldots 0.26\\
$1-1/ \delta $		&   2/3	 	&   0.8       & 0.7939(9)& 0.79(4)\\
$(1-\beta)/ \beta \delta $ &  1/3	& 0.3 $\pm$0.06 & 0.331(7) & 0.65(7)\\ 
$\alpha / \beta \delta $ &   0   &  -0.2 $\pm$0.1       & -0.13(3) & -0.07 \ldots +0.34\\
\end{tabular}\\[1ex]
\caption{\label{tab1}  Critical exponents of the three dimensional 
$O(4)$ model in mean field, in our approach with numerical errors (RG) 
and lattice calculations compared to lattice
QCD calculations with staggered quarks (QCD Lattice) (cf.~ref.~\cite{kana}).}
\end{center}
\end{table}

Because we know the free energy density of the system we can calculate
the specific heat directly , which is presented in the next section
and verify the above predictions by 
a determination of the critical exponent $\alpha$.  
We find numerically $\alpha = -0.39$ which is consistent with the 
scaling relation prediction and the $O(4)$ scaling is confirmed 
(cf.~fig.~\ref{fig8} and tab.~(\ref{tab1})). One should note though, that the
errors quoted for our results correspond to numerical errors but do not 
capture truncation errors.

An especially interesting point is to find the window for the
critical dynamics.
Further investigations away from the critical point
may give an indication how far away from $T_c$
the renormalization group improvement is important for the behaviour 
of the order parameter. 
In the superconducting phase transition only the
mean field behaviour is experimentally relevant. 
Figure~\ref{fig6} 
shows that in chiral QCD the critical behaviour influences a wide
range of temperatures, i.e.~from $T= 130$ MeV to the critical 
temperature $T_c$. 
In real QCD, however, the finite quark
masses spoil the second order phase transition. Also gluon effects will
change the behaviour of the pressure in comparison with the effective
linear $\sigma$-model. 
Our findings for the critical exponents show excellent agreement with
numerical simulation results of the lattice $O(4)$-model.

Numerical simulations of lattice QCD, however, indicate a 
critical behaviour of the order parameters 
which is  still ambiguous. The critical
index $\delta$ 
gives a strong hint for dimensional reduction.
For
the exponent $\beta$ a possible overlap exists. For the specific heat the 
QCD gluonic degrees of freedom may alter the behaviour significantly.
The treatment of light quarks and pions 
with large correlation lengths 
is  problematic on the lattice.


\subsection{Free energy at finite temperature}

Now we turn to the discussion of the equation of state for the 
linear $\sigma$-model
in the chiral limit. We emphasize this point, because
the physics of the model
calculation is clarified by the equation of state.  
If the model is complete in the low virtuality region,
the transmutation of quark-gluonic degrees of
freedom occurs  above the ultraviolet
scale. Is this assumption justified?

Considerable care has been devoted
to the treatment of critical fluctuations. In order to mark 
the progress achieved so far, the equation of state is the next
goal. Of course one does not expect to see free gluons
above $T_c$, but what kind of interaction effects among the quarks
do we have? Can they be compared to deviations of the lattice
results from the Stefan Boltzmann limit?

The most important thermodynamical tool for the analysis of a 
phase transition is the free energy, which is given as a logarithm 
of the partition function. The logarithm of the partition function is
equal to the effective action, defined in eq.~(\ref{potential}). Thus
the solution of the flow equations, (\ref{1flowtsym}) 
and (\ref{1flowtbroken}), yields the free energy density $f$ 
as function of
the temperature so that the pressure is given by
\ba
\label{defpressure}
P(T) & = & -v_{k \to 0} = -f\ .
\ea
The internal energy density is defined by 
\ba
\label{definternal}
u & = & \frac{d (\beta f)}{d \beta } \\
& = & - T^2 \frac{ d (\beta f)}{d T }\ ,
\ea 
with $\beta = 1/T$ and the entropy density by
\ba
\label{defentropy}
s & = & -\frac{d f}{d T}\ .
\ea 
These quantities are  connected by the relation $P = Ts -u$.

Finally, the specific heat is given by a further temperature derivative of
the free energy density
\ba
\label{defspecific}
c_V & = & -\beta^2 \frac{d u}{d \beta} \nonumber \\
&& = \frac{d u_k}{d T}\ .
\ea

In order to find the temperature derivative of the free energy density
one has to take into account the temperature 
dependence of
the quartic coupling $\lambda_k (T)$ and the VEV $\phi_k (T)$. 
In principal the thermodynamic functions can be obtained by numerically 
differentiating the $k=0$ value of the free energy density as a function
of temperature. In practice  the accuracy of this purely numerical 
differentiation is not sufficient, especially around the critical temperature. 
We therefore prefer to create for each phase 
a set of additional six evolution 
equations which contain the first and the second derivatives with respect 
to temperature of the original evolution equations, 
cf.~eqs.~(\ref{2flowtsym}),(\ref{3flowtsym}) for the symmetric phase and
eqs.~(\ref{2flowtbroken}),(\ref{3flowtbroken}) for the broken phase.
Thus for each temperature derivative one has to
solve three additional flow equations in this truncation schema, i.~e.~in
total we have 
nine highly, non-linear coupled flow equations.
Because they
are lengthy and are given by a straightforward 
calculation we renounce to quote the result here. 

\begin{figure}[ht!] 
\begin{minipage}[t]{0.5\linewidth}
	\centering	
	\includegraphics[height=6cm,width=7cm]{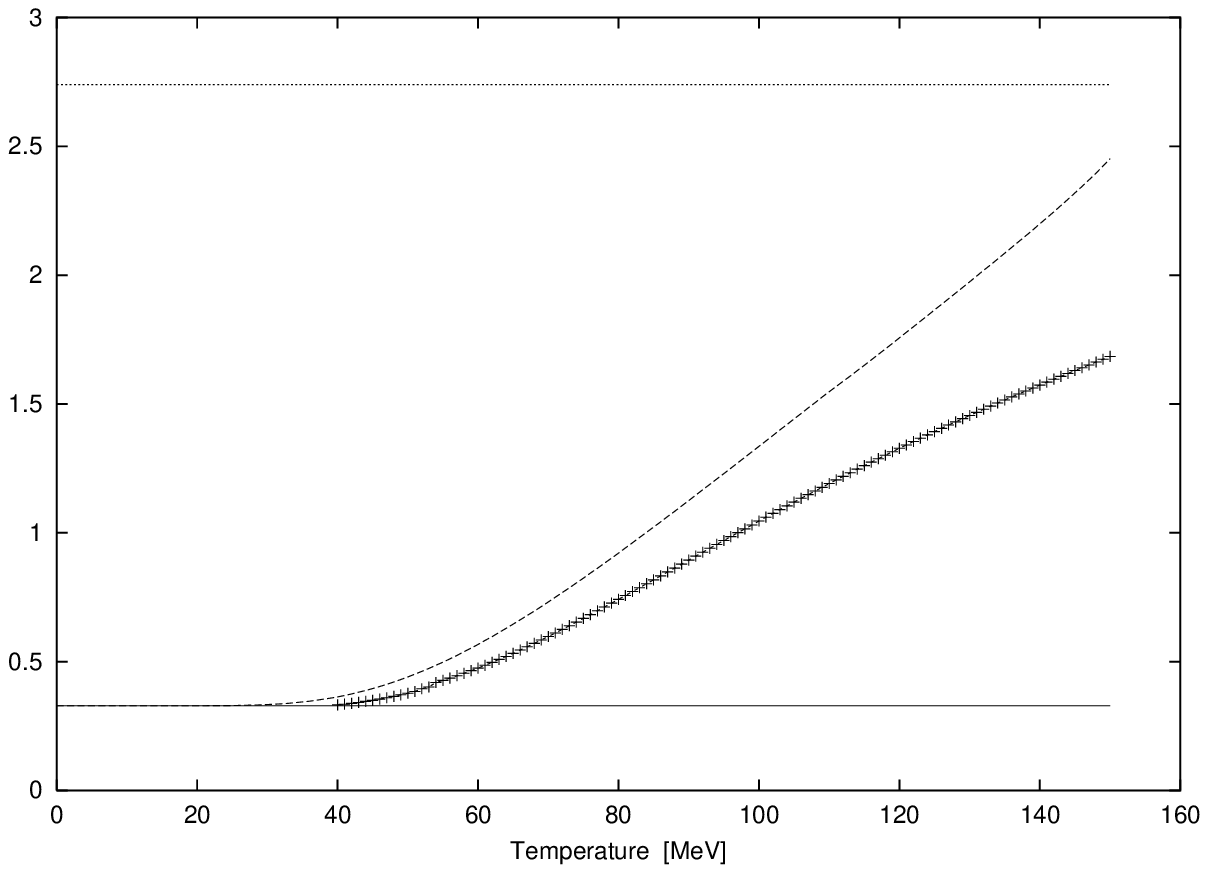}
	\caption{Pressure $P/T^4$} \label{figpressure}
\end{minipage}
\begin{minipage}[t]{0.5\linewidth}
	\centering	
	\includegraphics[height=6cm,width=7cm]{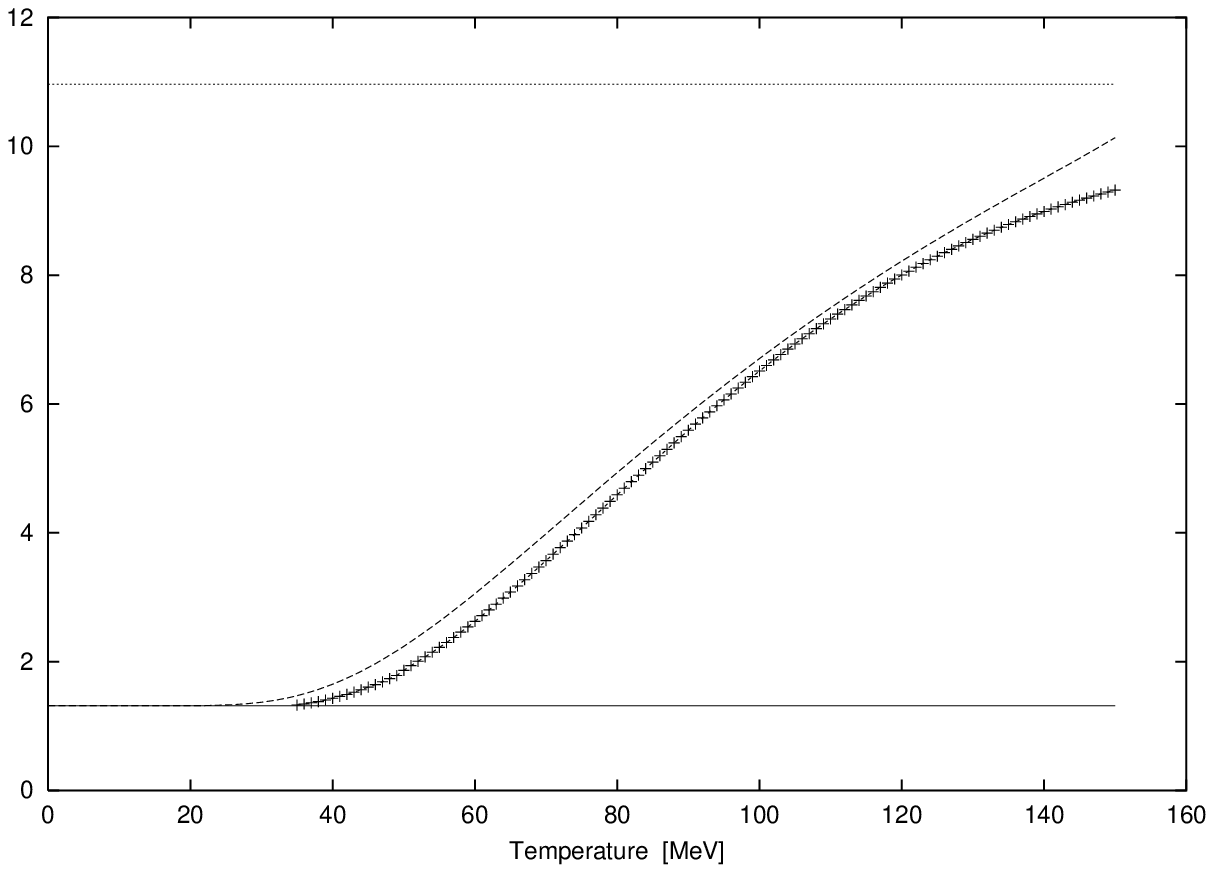}
	\caption{Entropy density $s/T^3$} \label{figentropy}
\end{minipage}
\end{figure} 
\begin{figure}[ht!] 
\begin{minipage}[t]{0.5\linewidth}
	\centering	
	\includegraphics[height=6cm,width=7cm]{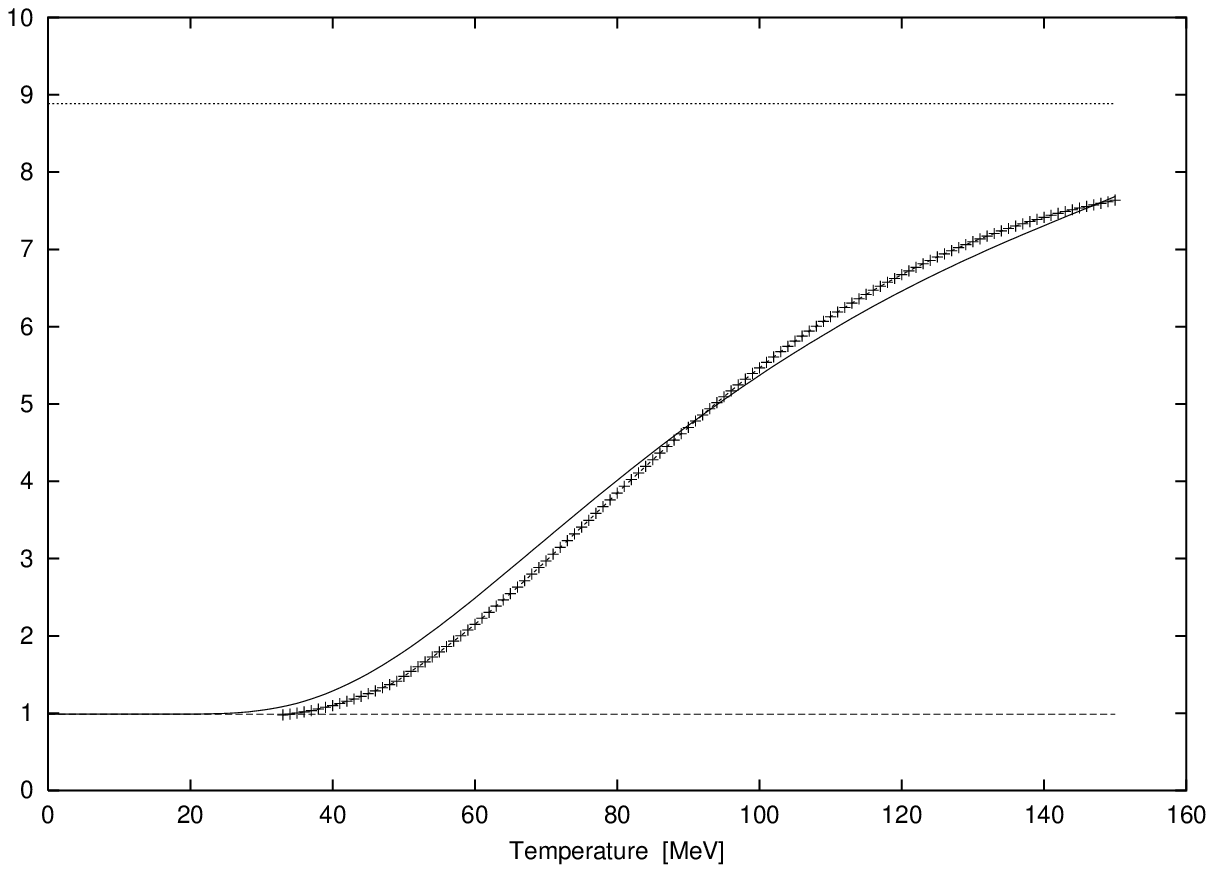}
	\caption{Internal energy density $u/T^4$} \label{figinternal}
\end{minipage}
\begin{minipage}[t]{0.5\linewidth}
	\centering	
	\includegraphics[height=6cm,width=7cm]{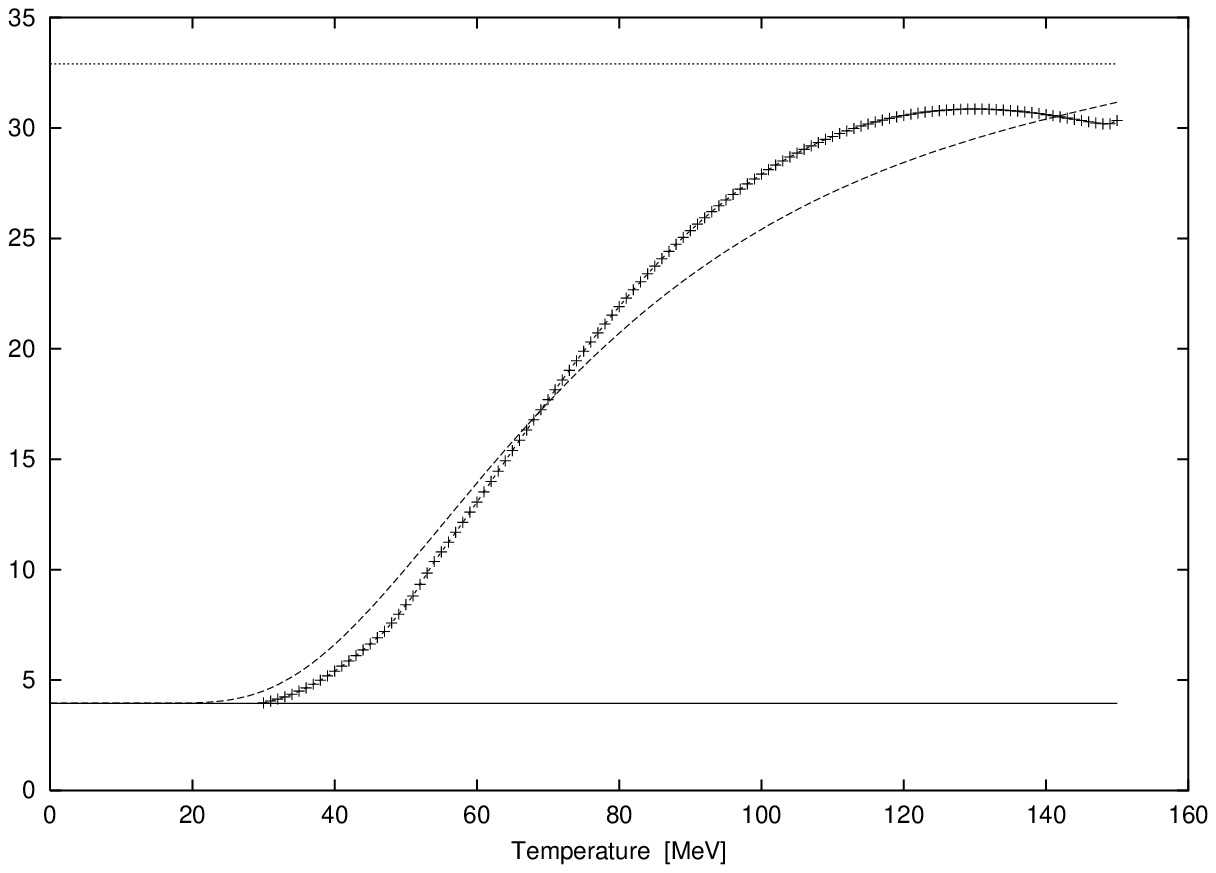}
	\caption{Specific heat $c/T^3$} \label{figheat}
\end{minipage}
\end{figure}

The procedure to solve these equations is the same as before.
The thermodynamic predictions are restricted to a temperature
window, which is dictated by the omission of the temperature dependence
of the initial values for the flow equations at the ultraviolet cutoff
scale (cf.~discussion sec.~\ref{seccritical}). From this point of view
a large scale to start the
evolution is desirable in order to reach predictive power at least up to
the critical temperature. For all calculated thermodynamic quantities we
choose a ultraviolet cutoff $\Lambda = 1.52$ GeV, which is close to the 
Landau pole of the quartic coupling in the chiral limit. 
Due to this Landau pole inherent 
in the flow equations and the necessity to satisfy certain 
phenomenological conditions discussed below 
a higher choice for the  
ultraviolet scale is not possible.
We also changed $f_\pi$ to $88$ MeV in the chiral limit, so that we obtain
for the finite pion mass case the measured pion decay constant
$f_\pi = 93$ MeV automatically (cf.~sec.~\ref{secfinitepionmass}). 
With the new ultraviolet cutoff we need
new initial values. We take $\lambda_\Lambda = 225$, which is larger
than the one used before with a smaller cutoff due to the growth of the
quartic coupling in the ultraviolet. We have to adjust the starting mass
to $m_\Lambda =1$ MeV, because for high cutoff with large $\lambda_k$ the
evolution equation of $m^2_k$ (cf.~eq.(\ref{2flowsym})) is governed by the
bosonic term proportional to $\lambda_k$ which first drives the mass up for
$\Delta k < 0$ before the fermion loop decreases $m_k$ again.

For the flow equations we find for the
critical temperature the smaller value
\ba
T_c = 149 \mbox{MeV}\ ,
\ea 
which follows from the smaller $f_\pi$, but is still much higher 
than the value found by the Wetterich group~\cite{wett}.

The initial values for the additional flow equations at the
ultraviolet scale $\Lambda$ are zero.
At the chiral symmetry breaking scale $k_{\chi SB} \approx 0.8$ 
GeV which is also a function of temperature the slopes of the 
flow equation for $v_k$ in the symmetric phase (\ref{1flowtsym}) 
and in the broken phase (\ref{1flowtbroken}) are identical. The 
evolution of the potential $v_k$ with respect to $k$ already stabilizes
at small but finite $k-$values ($k \approx 150$ MeV).

\begin{figure}[hbtp]
\unitlength1cm
\centering \leavevmode
\put(5.5,-0.5){Temperature [MeV]}
\includegraphics{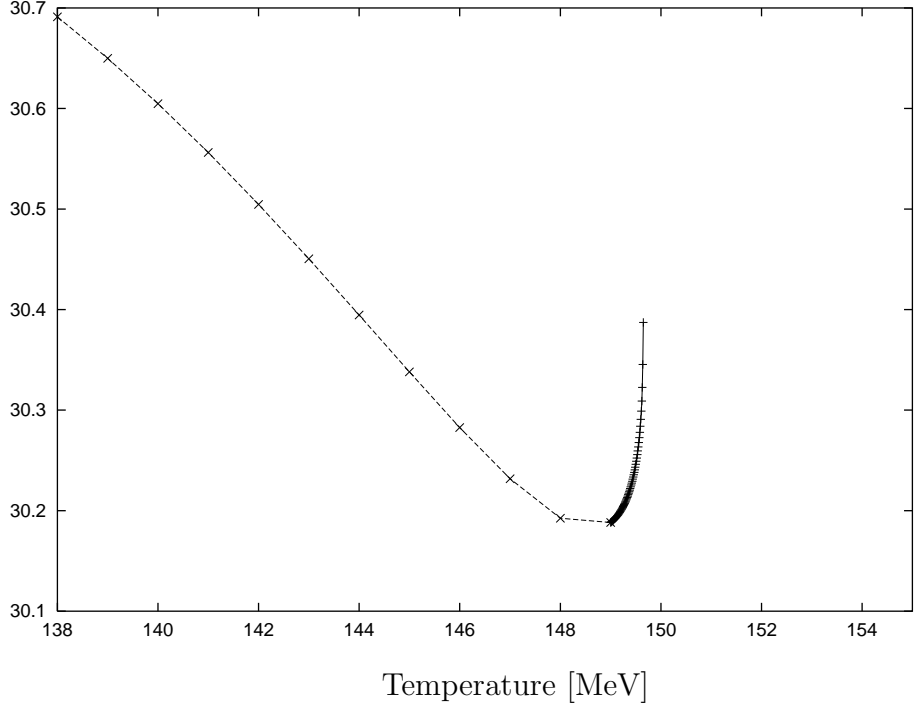}
\parbox{12cm}{
\caption{\label{figheatdetail}Specific heat $c/T^3$ in the 
	vicinity of the critical temperature.} }
\end{figure}

The result for the normalized pressure 
(cf.~eq.~(\ref{defpressure})), scaled with the temperature $P/T^4$, 
is plotted
in figure \ref{figpressure} and in figure \ref{figentropy} the 
scaled entropy density $s/T^3$ (crosses) is shown. 
In figure \ref{figinternal} the scaled internal energy density $u/T^4$ 
and in figure \ref{figheat} the scaled specific heat $c/T^3$ are
shown for the chiral limit.

In order to compare our results with a gas of
non-interacting  quarks, antiquarks  and pions,
we evaluate the corresponding thermodynamic functions
with the  temperature dependent masses evolved to $k \approx 0$
(dashed lines in the figures).

For small temperature the quarks are massive and therefore 
thermodynamically suppressed by their Boltzmann factor. 
The main contribution 
to all presented thermodynamic observables in that temperature 
limit stems from the massless pions.

We stop the flow equation at a small infinitesimally   
finite scale $k$ and obtain a finite sigma meson mass due to a finite
quartic coupling. We have checked that there is no influence of a sigma 
degree of freedom to the thermodynamic functions at finite temperature, 
because the relevant quantity for decoupling is the mass squared 
over $k^2$.

Due to the linear evolution 
of the quartic
coupling with $k$ for $k \ll T$ 
which tends to zero for $k=0$ in the chiral limit, 
the sigma meson decouples 
from the low momentum mode thermodynamics and we obtain
the expected Stefan-Boltzmann limit
of a massless pion gas for small temperatures.

The respective thermodynamic functions 
for three pionic degrees of freedom are given by
\ba
P/T^4 & = & 3 \frac{\pi^2}{90} \approx 0.33\\
u/T^4 & = & 3 P/T^4 = 3\frac{\pi^2}{30} \approx 0.99\\
s/T^3 & = & 4 P/T^4 = 12 \frac{\pi^2}{90} \approx 1.32\\
c/T^3 & = & 4 \frac{\pi^2}{10} \approx 3.95
\ea

In each figure \ref{figpressure}-\ref{figheat} 
the lower horizontal lines represent the pion gas limit. 
With increasing temperature all observables except the pressure
follow roughly the ideal gas behaviour. 
At the critical temperature the quarks are massless
and the Stefan-Boltzmann
limit for a massless quark-antiquark and massless pion-sigma meson gas
is depicted by the upper horizontal line in the figures.
The thermodynamic functions for a gas with
$N_c N_f$ massless quarks can be calculated analytically:
\ba
P/T^4 & = & N_c N_f \frac{28\pi^2}{720} \approx 2.30\\
u/T^4 & = & 3 P/T^4 = N_c N_f \frac{28 \pi^2}{240} \approx 6.9\\
s/T^3 & = & 4 P/T^4 = N_c N_f  \frac{112\pi^2}{720} \approx 9.21\\
c/T^3 & = & N_c N_f \frac{28\pi^2}{60} \approx 27.63
\ea

For the numerical calculations at finite temperature one can not
simplify the computer effort. In ref. \cite{wett} the authors reduced 
the complicated and time intensive  calculations by
substituting the simpler zero temperature flow equations for
the finite
temperature flow equations as long as $k \ge 10 T$. For
smaller scales they used the finite temperature flow equations. This is
certainly a good approximation for the 
quantities considered there. 
But the low temperature equation of state 
would not be described correctly in such an 
approximation. 
We therefore use during the whole evolution the  finite
temperature flow equations and optimize the convergence of the
Matsubara series by an generalized $\Theta$-function transformation,
which is
described in the appendix. In practice
we take more than $N=4000$ Matsubara frequencies into account in the
low temperature region ($k \gg T$) in order to adequately
solve the equations over the hole scale range.  
For higher temperatures ($k \ll T$) only a few Matsubara modes are important.

On the other hand 
our ultraviolet scale might be identified with the
scale \cite{wett}, where the meson bound states appear. 
Due to the neglect of the 
wave function renormalization 
we do not suppress the kinetic terms of the mesons at the compositeness scale.
This has an important consequence: Fixing the 
Yukawa coupling $g$ at the compositeness scale to its infrared
value which yields a finite constituent quark mass of roughly
$m_q \approx $ 300 MeV, we 
do not encounter the partial approximate 
infrared fixed point, found in \cite{wett}, 
which would weaken the dependence of the 
infrared physics on the chosen initial values.  
In our case we can  choose a high ultraviolet scale
and are nevertheless able to adjust the  infrared quantities like 
e.~g.~the pion
decay constant.

To improve the finite temperature results we use a so-called  
generalized $\Theta$-function 
transformation which accelerates the convergence 
of the Matsubara sums leading to  modified Bessel function $K_\nu
(x)$. This reformulation has the pleasant side effect to
separate analytically the thermal fluctuations from the $T=0$
fluctuations. The right hand sides of the flow equations
split into the pure zero temperature flow equation part and a finite
temperature part. 
We are aware that the separation of these terms 
does not decouple the quantum  from the thermal fluctuations because
of the nonlinearity of the evolution.

This property, however,  allows a clean normalization:
neglecting  the finite temperature part of the corresponding flow
equation yields the correct zero temperature contribution.

In our approximation of neglecting the scale and temperature dependence 
of the Yukawa coupling as well as the quark and meson wave function 
renormalization,
the constituent
mass follows the temperature behaviour of
the order parameter $\phi_k (T)$. In the chiral limit above the
critical temperature of $T_c \approx 150$ MeV the quark masses are
identically zero and the mesonic masses are degenerate and increase with
temperature because the mesonic coupling also increases with
temperature in the symmetric phase. For finite pion masses this
behaviour is smoother, but  qualitatively similar and will be
discussed in the next section.  


\section{Equation of state for finite pion mass}
\label{secfinitepionmass}

It is straightforward to  derive thermodynamic quantities for
finite pion masses by introducing an explicit symmetry breaking term 
$c$ in the potential. 
The explicit symmetry breaking constant $c$ is kept fixed and is
proportional to
the current quark mass. 

We find the following mass identifications: 
\ba \label{pionmass}
\mbox{pions:}\qquad m^2_{\pi,k} (T) & = & c/\phi_k (T)\\
\mbox{sigma:}\qquad m^2_{\sigma,k} (T) & = & c/\phi_k (T) + 
2\lambda_k (T) \phi_k^2 (T)\\
\label{quarkmass}
\mbox{quarks:}\qquad M_{q,k} (T) & = & g \phi_k (T)
\ea
It is 
not clear if the $\sigma$-mass, given above, is actually the physical
$\sigma$-mass and necessarily a pole in the two-point Greens function.
We follow ref.~\cite{wett} where this expression is also denoted as the
$\sigma$-mass and corresponds to the radial mode of the $O(4)$ model. 

\begin{figure}[hbtp]
\unitlength1cm
\centering \leavevmode
\put(-1.0,7.5){[MeV]}
\put(2.0,5.5){$m_\sigma$}
\put(2.0,2.0){$m_\pi$}
\put(5.0,-0.5){Temperature [MeV]}
\includegraphics{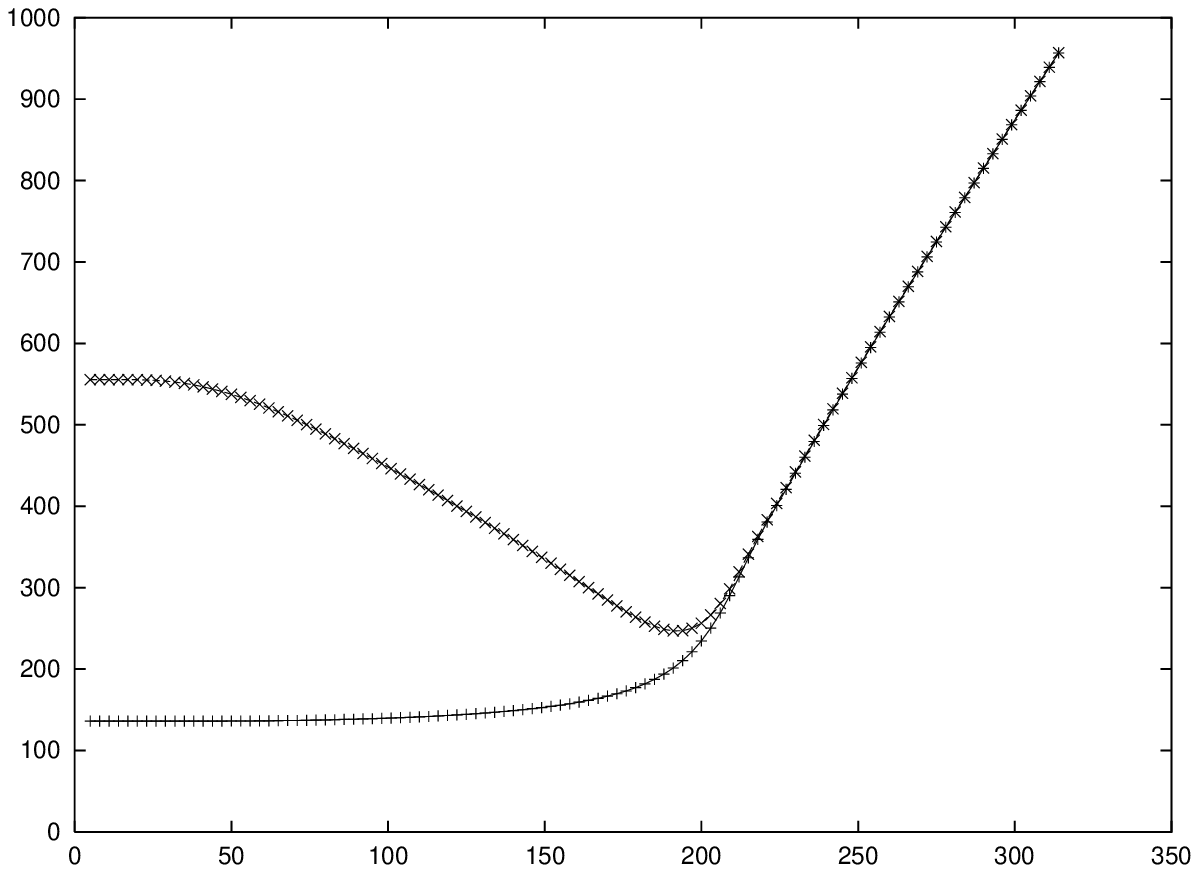}
\caption{\label{fig13} Mesonic masses versus temperature.}
\end{figure}

We fix the constant $c$ via the pion mass ($m_\pi (T=0) = 135$ MeV) to $c =
1.7\cdot 10^6$ MeV$^3$, which results in a sigma-mass of about 550
MeV at vanishing temperature.
In reference \cite{wett} the sigma mass
is considerably smaller.  
For finite pion masses the two sets of flow equations for each phase in the 
chiral limit merge together to one set of flow equations.
The constituent quark mass follows the
order parameter and never becomes zero in the symmetric phase for high 
temperature due to a smooth crossover between both phases.
The evolution equations for finite pion masses have been studied 
for two different cutoffs $\Lambda=1.2$ GeV and $\Lambda=2.0$ GeV.
Due to the symmetry breaking term in the potential the initial values
are not so sensitive to a change of cutoff. The respective initial
values are
for
$\Lambda = 1.2$ GeV $\lambda = 115$, 
$m = 1$ MeV and
$\lambda = 430$, 
$m = 1$ MeV for  
$\Lambda = 2$ GeV.
In fig.~\ref{fig13} the resulting mesonic masses are shown as function of
temperature. The scalar meson mass drops relatively quickly with 
temperature while the pseudoscalar masses are almost temperature 
independent in the broken phase. We think that a wave function 
renormalization correction for the scalar fields 
does not change this situation dramatically at least in the broken phase.

\begin{figure}[hbt]
\unitlength1cm
\centering \leavevmode
\begin{picture}(15,9)(-1,-0.5)
\put(5.0,-0.5){Temperature [MeV]}
\put(-0.9,8.0){$\phi_{k=0}$}
\put(-1.0,7.2){[MeV]}
\includegraphics{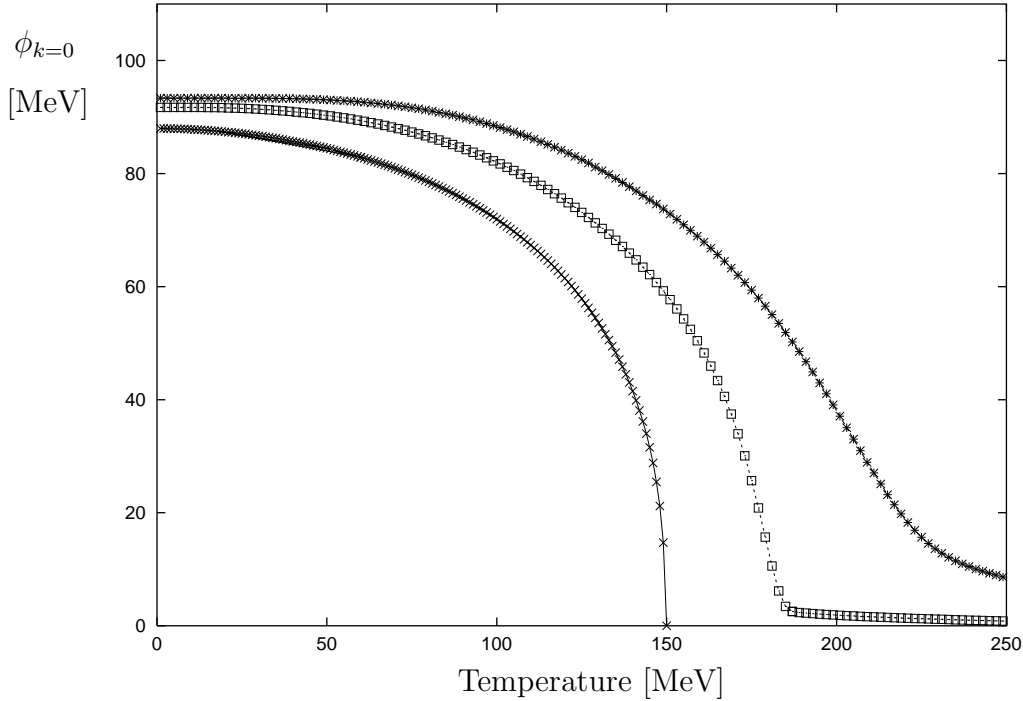}
\end{picture}
\parbox{12cm}
{\caption{\label{fig14} The order parameter $\phi_{k=0}$ 
for different pion masses. 
(Lower curve: chiral limit; middle curve: $m_\pi \approx 45$ MeV; 
upper curve: $m_\pi \approx 135$ MeV).}}
\end{figure}

In general the derivative of these masses with respect to temperature
is independent
of the choice of the initial UV cutoff.  
The increase of the masses with temperature in the
restored phase, however,  depends slightly on the choice of the UV starting
cutoff. We consider this as  an artefact of the omission of  
wave function renormalization, which would influence
the temperature increase of the quartic coupling in the restored phase.

In fig.~\ref{fig14} we compare the order parameter for different pion
masses. The lower curve represents the chiral limit where one sees
a second order phase transition. This transition is smeared out if
one switches to finite pion masses. The middle curve is achieved for
a pion mass of roughly $45$ MeV and the upper curve shows the realistic
case. Note that the order parameter at zero temperature increases with
the pion mass automatically. If one defines a pseudocritical temperature
analog to ref.~\cite{wett} as the inflection point of the order parameter
one observes an increase of the critical temperature with the pion mass
from $T_c \approx 150$ MeV for the chiral limit up to $'T_c' \approx 190$
MeV for a pion mass $m_\pi = 135$ MeV. Due to the finite symmetry breaking
term in the potential the order parameter never vanishes at high temperature. 

Above the critical temperature of the chiral transition about $200$ MeV 
we see a parity doubling of the scalar and pseudoscalar particles, which 
is a signal of chiral restoration and the masses for the mesons 
rise proportional to temperature due to the increase of the  effective
mass term with temperature and the symmetry breaking term.
This behaviour is similar to the behaviour of meson correlation functions
for 
massless free quarks above $T_c$ on the lattice \cite{elet}.

The hadronic screening masses at high 
temperature are controlled by the lowest Matsubara 
frequency of quarks and increase therefore like $2\pi T$. 
The physics in our case is probably more related to the meson dynamics.

\begin{figure}[hbt!]
\unitlength1cm
\centering \leavevmode
\put(5.0,-0.5){Temperature [MeV]}
\put(2.0,7.5){\bf pressure $p/T^4$}
\put(7.5,8.0){ideal quark + meson gas}
\put(9.5,4.7){$\Lambda = 2$ GeV}
\put(7.5,3.0){$\Lambda = 1.2$ GeV} 
\put(6.0,1.7){ideal pion gas}
\includegraphics{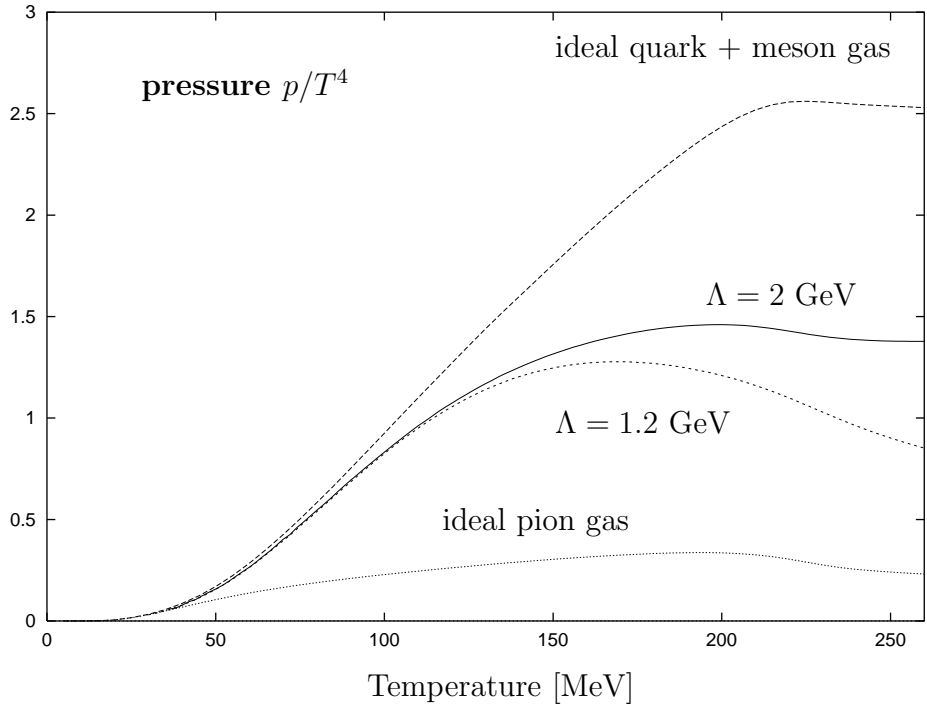}
\parbox{12cm}
{\caption{\label{pressure.vergleich} Scaled pressure 
  $(P(T)-P(T=0))/T^4$ for different systems as function of temperature 
  (see text for details).}}
\end{figure}

In the remaining section we focus on the thermodynamic.
In general the following physical 
scenario becomes evident: For low temperature the quarks and
mesons are massive, while in the partition function with 
unbroken chiral symmetry the pions remain 
massless\footnote{ In this case the so-called 
  sigma meson mass, which is proportional to the quartic coupling tends 
  to zero in the IR region.}.
In the chiral limit for low temperature one expects and indeed
obtains  that the
physical observables are  dominated by a massless non-interacting 
pion gas and 
that the Stefan-Boltzmann limit is saturated. On the other hand,
for high temperature in the thermally restored chirally symmetric 
phase the quarks are exactly massless while both,
the pions and the sigma mesons, have finite (degenerated) 
masses due to the restoration of chiral $SU(2) \times
SU(2)$ symmetry. Thus all thermodynamic
quantities at high temperature are dominated by the massless quarks with 
$N = 4N_c N_f$ degrees of freedom modified by their interaction.

\begin{figure}[hbt]
\unitlength1cm
\centering \leavevmode
\put(5.0,-0.5){Temperature [MeV]}
\put(1.5,7.5){\bf energy density $u/T^4$}
\put(9.0,5.2){$\Lambda = 2$ GeV}
\put(8.0,2.8){$\Lambda = 1.2$ GeV} 
\put(6.0,1.5){ideal pion gas}
\put(7.5,8.0){ideal quark + meson gas}
\includegraphics{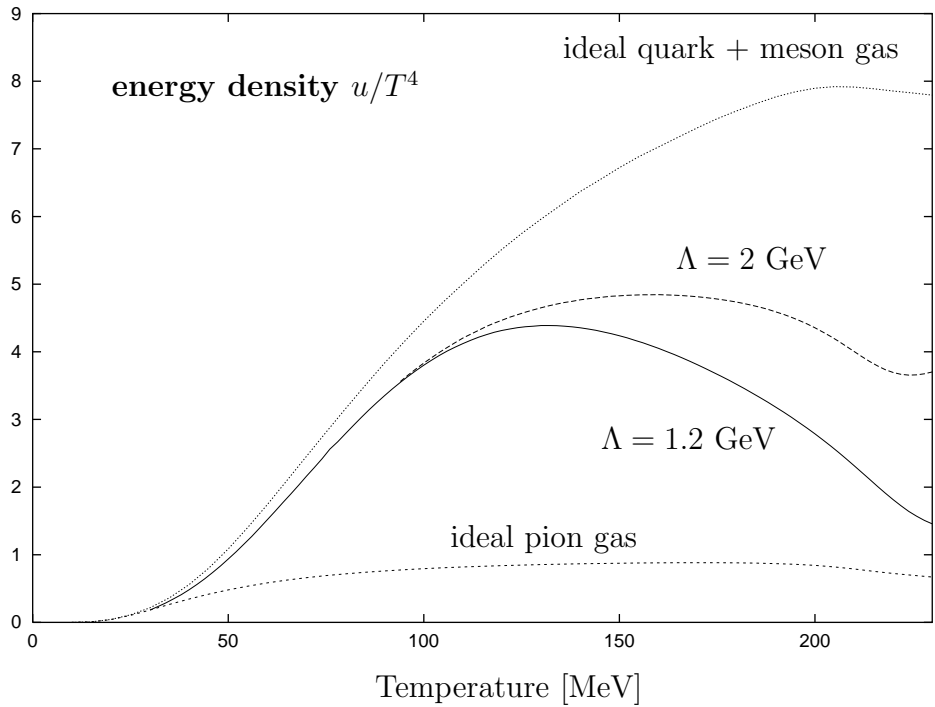}
\parbox{12cm}
{\caption{\label{internal.vergleich} Scaled energy density 
  $u/T^4$ for different systems as function of temperature 
  (see text for details).}}
\end{figure}

\begin{figure}[hbt]
\unitlength1cm
\centering \leavevmode
\put(5.0,-0.5){Temperature [MeV]}
\put(2.0,7.5){\bf entropy density $s/T^3$}
\put(7.5,8.0){ideal quark + meson gas}
\put(9.5,5.1){$\Lambda = 2$ GeV}
\put(8.0,2.7){$\Lambda = 1.2$ GeV} 
\put(6.0,1.7){ideal pion gas}
\includegraphics{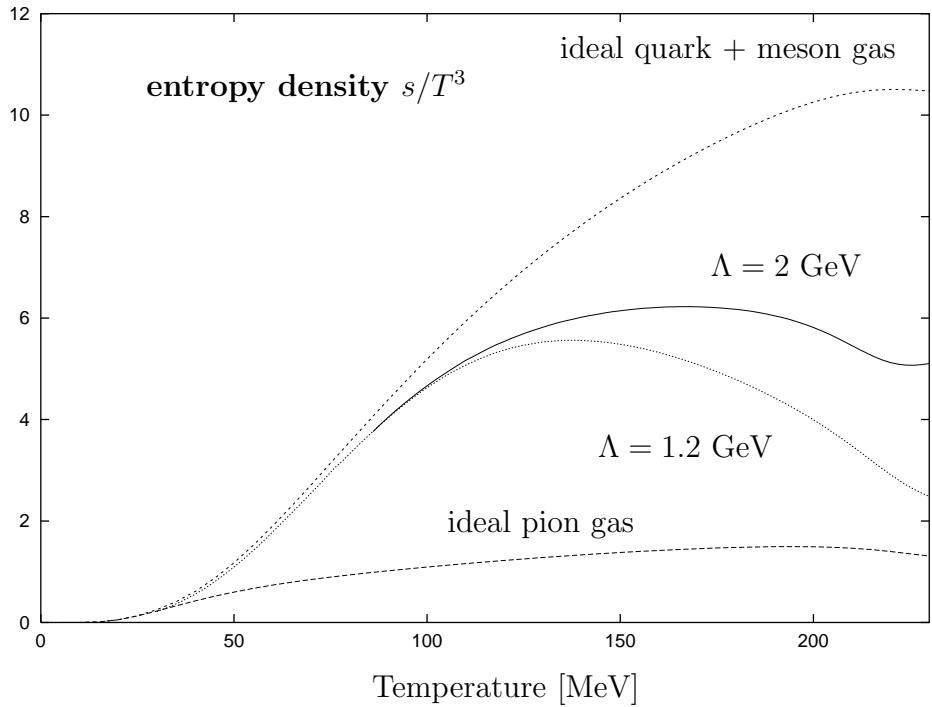}
\parbox{12cm}{
\caption{\label{entropy.vergleich}  Scaled entropy density $s/T^3$
  versus temperature for different systems (see text for details).}}
\end{figure}

A similar but not identical situation holds for  finite quark masses.
The low temperature gas is hadronic made out of pions mostly,
the high temperature plasma is a quark plasma.
At the critical point  the sigma mass is small.
Therefore, except near the critical temperature the physical 
sigma meson contribution to the
thermodynamic functions for low and high temperature can nearly be neglected.
In the following figures we show the pressure
$P/T^4$ (cf.~fig.~\ref{pressure.vergleich}) 
and the scaled internal energy density $u/T^4$
(fig.~\ref{internal.vergleich}) and the scaled
entropy density $s/T^3$ (fig.~\ref{entropy.vergleich}). 
In all figures we compare the results 
with a non-interacting quark-meson gas (uppermost curve) 
and also with a ideal pion gas
(lowest curve).

For finite masses the pressure
for small temperature is exponentially suppressed as expected.     
At $T \approx 200$ MeV the total pressure  
caused by interacting pions, sigma
and constituent quarks reaches the maximum at $(P(T)-P(T=0))/T^4 = 1.46$, 
decreases in the following a bit 
and settles down around $T \approx 300$ MeV at $1.37$ (see
fig.~\ref{pressure.vergleich} ).

A pressure due to the quarks exists also for small temperature, which
is a consequence of the fact
that the quarks are not confined in the model. For zero temperature
there is no pressure because the quarks are massive but with
increasing temperature, the pressure rises 
and at temperatures around $T \approx
70$ MeV the quark pressure  is comparable with the mesonic
pressure. In fig.~\ref{pressure.vergleich} the lowermost curve
shows the pressure  of $N=3$ pions.

With increasing temperature
the quark contribution 
takes over and the total pressure  saturates at $\sim 55
\%$ of the Stefan-Boltzmann 
limit for $T > 250$ MeV. This result is comparable with an NJL model
calculation for small temperature 
using a RPA approximation including fluctuations. Even below  the 
critical temperature 
a
strong cutoff dependence enters into the results of previous 
calculations \cite{zhua}.     

At very high temperatures $T>180$ MeV 
our model is no longer adequate: the pressure
depends on the ultraviolet cutoff which suppresses high 
momenta.
Therefore all non-universal thermodynamical quantities have a tendency
to be too small at too high temperatures.
In order to demonstrate this property we have plotted
two curves in the figures. The lower line displays the 
pressure  for an ultraviolet cutoff $\Lambda = 1.2$ GeV while the upper
curve is calculated with $\Lambda = 2.0$ GeV. One sees nicely which
temperatures can be described for a given
cutoff. Both curves coincide up to
a temperature of $T \approx 120$ MeV,  the lower curve
turns off for higher temperature. Indeed it
should be
valid only up to $T \approx 120$ MeV. We have checked that the
results in the shown temperature region are stable with respect to a further 
increase of the ultraviolet cutoff beyond $\Lambda = 2.0$ GeV.

The uppermost curve in 
figure~\ref{pressure.vergleich} displays the total contribution
of a non-interacting gas of quarks and mesons
calculated with temperature  
dependent masses, which are extracted out of the flow equations.
It reaches almost the
Stefan-Boltzmann limit ($P/T^4 = 2.7$) around $T \approx 230$ MeV due to the
decrease of the vacuum expectation value and therewith decreasing
quark masses. On the other hand, the pions and
sigma-meson pressure increases ($m_\sigma$ decreases) up to $T \approx 
200$ MeV and settles down 
in the symmetric phase on a constant value. This reflects the fact,
that the meson
masses increase proportional to the temperature and the ratio
mass over temperature is almost constant. 
The pressure from the flow equation is considerably smaller than this
limiting Stefan Boltzmann Plasma. 

In fig.~\ref{internal.vergleich} the energy density, scaled by $T^4$,
is shown as function of the temperature. The interpretation of the
results and notation of this figure is
analogous to the pressure  calculation. In the chiral limit at
low temperatures we
observe again the Stefan-Boltzmann limit $u/T^4 = \pi^2/10$ 
of the massless pion modes ($N=3$). For finite pion mass
at high temperatures the dependence on 
the ultraviolet cutoff is slightly 
stronger compared to the 
pressure  calculation.

The entropy density, scaled by $T^3$, is plotted as function of
temperature in fig.~\ref{entropy.vergleich} and is obtained as a
derivative with respect to the temperature of the free energy density,
$s = - \partial f / \partial T$.
The low and high temperature features of this quantity can be
understood in a similar fashion as the pressure and
energy density. For finite pion masses the entropy density is
continuous in the vicinity of the chiral critical temperature ($T_c
\approx 150 $ MeV) and reaches  $60 \%$ of the Stefan-Boltzmann
limit around these temperatures, which
equals to $s/T^3 = 8N_c\cdot 31\pi^2/720 + 16 \pi^2/90 =
11.95$. Due to finite meson masses which increase proportionally to
temperature in the symmetric regime, the bosonic contribution to all
thermodynamic quantities is strong suppressed. 
All shown thermodynamic quantities
are reliable in a temperature window of $0 \le T \le 180$
MeV for an ultraviolet cutoff $\Lambda$ of  $2.0$ GeV.


\section{Conclusions}
\label{sec5}
We have presented a new method to solve field theoretic problems
at finite temperature. It is based on a nonperturbative application of
the renormalization group. By successively integrating out modes in 
Euclidean momentum shells we can handle the nonlinear field couplings 
in a nonperturbative way,  
especially when long wavelength modes 
generate singular infrared behaviour near the critical temperature.

We apply a cutoff function 
in  Schwinger 
proper time which  combines a virtuality- or mass- like exponential cutoff 
with a polynomial which makes the calculation 
of the derivative of the effective potential with respect to
the evolution scale insensitive to the ultraviolet
modes. 
This reflects the modern view of renormalization theory, all 
field theories
are effective theories with an ultraviolet cutoff and we always follow
their infrared behaviour.  
The evolution equations are finite and well behaved, which
a simple mass type cutoff would not guarantee. The specifically 
chosen heat kernel cutoff
function has the additional advantage that all resulting evolution equations
can be expressed in a  very transparent and analytic form. 
In the one loop approximation 
propagator like functions for the individual particles appear with
denominators which contain the 
respective masses depending on the effective potential. The 
renormalization group allows to improve the one loop approximation to sum
in an optimal way the infrared singular fluctuations due to massless 
boson modes.

In this paper we have in addition truncated the effective
potential to its lowest terms in the boson fields compatible with chiral
symmetry. This  is a simplification which is not necessary and can be
improved \cite{papp},\cite{aoki}. Also the consideration of coupling
constant and more importantly wave function renormalization 
evolution equations is planned. 
The inclusion of vector and axial-vector mesons 
at finite temperature to study their mixing
is also possible within this method and work in 
this direction is in progress~\cite{dabj}.

This new method has been applied to the finite temperature phase
transition from a constituent quark gas interacting with mesons to a
massless quark phase in the framework of the  linear sigma model. The 
underlying assumption of a reduction of the QCD degrees of freedom to 
an effective hybrid quark meson field theory at an ultraviolet scale of
about $1$ GeV has been analyzed via the resulting equation of state. 
Of course, qualitatively the pressure  is always dominated by the massless 
modes which are the  pions at low temperature and the quarks at high
temperature. But our calculation has highlighted  finer features of
the finite temperature  behaviour.

The evolution equations decouple the quarks from
further evolution when $T/k$ becomes too large. This does not mean,
however, that the quarks do not contribute to the pressure below the
chiral phase transition. The renormalization group technique cannot
mimic confinement when  the field theoretic model does not contain 
this dynamical aspect. The equation of state at high temperatures shows
considerable deviations from a free system of massless quarks.
In our interpretation of the chiral phase transition
long wavelength fermion modes really have
zero masses at $T_c$. Quarks with moderate or higher virtualities are
still massive and only at temperatures considerably higher than $T_c$
the whole virtuality spectrum of modes is massless.

The real strength of the  renormalization group lies  near 
the critical temperature where the $\sigma$ modes become massless.
Here the linear sigma model  with quarks is entirely dominated by the
bosonic sector. The evolution equations
reproduce well the universal critical behaviour of the model. We have 
calculated the critical indices  $\beta$ and $\alpha$ which 
are in good agreement with  lattice simulations of the $O(4)$
theory. The missing wavefunction renormalization correction is 
known to be small. The evolution equations
at finite temperature contain the physics of dimensional reduction:
For $T/k >0.3$ the bosonic loop contributions reduce to their
zero Matsubara frequency part, i.e. they become
equivalent three  dimensional loops with effective couplings 
which have acquired temperature factors. The fermions decouple 
in this region of $T/k$. The differences of some of the critical indices
to the mean field behaviour are small, i.e.  $\beta=0.4$
compared to $\beta_{mf}=0.5$. This does not mean, however, that the
region around $T_c$ where the critical dynamics is important is small
in the equation of state.

The phenomenological relevance of the
critical behaviour is more limited because of the finite quark mass
which breaks chiral symmetry explicitly and smoothes out all remarkable
features of the $O(4)$ behaviour. In that respect the chiral dynamics
is mostly important as a theoretical benchmark. It remains to be seen
how far the critical $O(4)$ behaviour is seen in  full QCD simulations.

The evolution equations have the unique property to unify 
different aspects of the hadronic system which conceptually and
experimentally are widely separated. The zero temperature evolution
gives a view of the field theoretic system with different resolution.
The virtuality cutoff $k$ is intimately related to the resolution
of a probe  with which we may study the system 
e.g.~with photons in deep inelastic
scattering. At high resolution we see a massless partonic system,
at low resolution we encounter massive constituent quarks. There are not
separate hadronic theories for different length scales: a high energy,
an intermediate energy and a low energy theory, ideally they all
evolve smoothly into each other. The challenge of hadronic physics
lies in the fact that the important degrees of freedom transmute in
a  strongly coupled theory, very much different from a weakly interacting
theory. Studying the hadronic system at finite temperature and/or finite
density allows to emphasize different modes by varying the external
parameter and thereby showing the transition of the dynamics explicitly.
The evolution equations unify these different aspects 
of QCD which when studied separately look like
rather independent  developments
in hadronic and nuclear physics.

\subsection*{Acknowledgments}
We would like to thank C.~Wetterich, D.-U.~Jungnickel and J.~Berges for their 
introduction  to the  subject. 
B.J.S. expresses his gratitude to J.~Wambach and D.-U.~Jungnickel for 
numerous enlightening discussions on large parts of this work. 
He also would like to
thank O.~Bohr and Z.~Aouissat for long and valuable discussions.   
H.J.P.~especially would like to 
acknowledge discussions with N.~Tetradis who has triggered this work.

\begin{appendix}

\section{Matsubara Series:}
\label{appmat}
In this appendix we summarize the low temperature limit of all
in this work used
Matsubara summations. 
They can be found by a comparison of the finite temperature flow equations
with the zero temperature equations. The expression of
eq.~(\ref{thresholdlimit})
can be generalized by
\ba
x \sum_{n = - \infty}^\infty \frac{1}{(1 + \tilde{\omega}^2_n + 
\tilde{y}^2)^\alpha}
 & \stackrel{\D x \to 0}{\longrightarrow} 
& \frac{2 \cdot 4 \ldots (2\alpha -3)}{1 \cdot 3 \cdot 5 \ldots
(2 \alpha -2)}\frac{1}{\pi} \frac{1}{(1 + \tilde{y}^2)^{(2\alpha -1)/2}}
\ea
with  $x := T/k$, $\tilde{\omega}^2_n := {\omega}^2_n/k^2$
and $\tilde{y}^2 := y^2/k^2$ and 
$\alpha = \frac{3}{2},\frac{5}{2}\ldots$ being a fractional power.
We set $(2\alpha -3 ) =: 1$ for $\alpha = \frac{3}{2}$.

For the derivatives with respect to temperature of the respective 
Matsubara sums we  find for the low temperature limit
($l =1$ describes the first derivative and
$l=2$ the second one) 
\ba
x^{2l +1}  \sum_{n = - \infty}^\infty 
\frac{(2n)^{2l}}{(1 + \tilde{\omega}^2_n + \tilde{y^2})^\alpha}\ & 
\stackrel{\D x \to 0}{\longrightarrow} & \\
&&\hspace{-5.5cm}
\frac{2l -1}{(2 \pi^2)^l (\alpha -1)(\alpha -2) \ldots (\alpha -l) }
\frac{2 \cdot 4 \ldots (2\alpha -3 - 2l)}{1 \cdot 3 \cdot 5 \ldots
(2 \alpha -2 - 2l)}\frac{1}{\pi} 
\frac{1}{(1 + \tilde{y}^2)^{ (2\alpha -1 -2l)/2}}\nonumber\ .
\ea
If one replaces $n \to n + \frac{1}{2}$ in the last equation
one gets the corresponding
fermionic Matsubara sums 
with the same low temperature limit. 
In the low temperature limit the difference between the fermionic and
bosonic Matsubara sums
vanishes. 

\section{Generalized $\Theta$-function transformation:}
\label{apptheta}
In order to accelerate the convergence of the Matsubara sums appearing 
in the flow equations it is useful to apply a generalized
$\Theta$-function transformation. 
This transformation is based on the following equations:

For bosonic Matsubara modes and positive integer $p$ 
we obtain\footnote{
A proof can be found in \cite{phde}.}
\ba\label{a4}
\sum_{n=-\infty}^\infty \left( \gamma n^2 \right)^p e^{-\gamma n^2}
& = & \sqrt{\frac{\pi}{\gamma}} \frac{(2p-1)!!}{2^p}
\left[ 1 + 2 \sum_{n=1}^\infty e^{-\pi^2 n^2/\gamma} \sum_{l=0}^p
\left( \frac{-\pi^2 n^2}{\gamma} \right)^l C^l_p \right]\qquad
\ea
with $(2p-1)!! = 1 \cdot 3 \ldots \cdot (2p-1)$ and
\bano 
C^l_p & = & \frac{2^{2l} p!}{(2l)! (p-l)!}\ ,
\eano
and for the fermionic case \cite{bosc} 
\bano
\sum_{n=-\infty}^\infty \left( \gamma (n+1/2)^2 \right)^p 
e^{-\gamma (n + 1/2 )^2}
& = &
\eano
\be
 \sqrt{\frac{\pi}{\gamma}} \frac{(2p-1)!!}{2^p}
\left[ 1 + 2 \sum_{n=1}^\infty (-1)^n e^{-\pi^2 n^2/\gamma} \sum_{l=0}^p
\left( \frac{-\pi^2 n^2}{\gamma} \right)^l C^l_p \right]\qquad\ 
\ee

This transformation can be applied to the finite temperature expression 
of the effective potential eq.~(\ref{potential2}) and the proper time
integration can be performed analytically 
resulting in modified Bessel functions $K_\nu (x)$. 

With the following 
convenient abbreviations
 ($B = $ bosons and
 $F = $ fermions)
 \ba
 h_\pi  &:=&  \frac{1}{1+c/(k^2 \phi_k)} \nonumber\\
 h_\sigma & :=&  \frac{1}{1+(c/\phi_k + 2\lambda_k
   \phi_k^2)/k^2}\nonumber\\
 h_q & := & \frac{1}{1+ g^2 \phi_k^2 /k^2}
 \ea
 and 
 \ba
 B_\nu (x_m) & := & h_m^{\nu/2} \sum_{n=1}^\infty n^\nu K_\nu (n x_m) 
 \qquad\mbox{$m$ = $\sigma$, $\pi$}\nonumber\\
 F_\nu (x_q) & := &  h_q^{\nu/2} \sum_{n=1}^\infty (-1)^n n^\nu K_\nu (n x_q)
 \qquad\mbox{$\nu$ : integer}\nonumber
 \ea 
 \ba
 x_\pi :=  \frac{k}{T} h_\pi^{-1/2}\quad
 & x_\sigma :=   \frac{\D k}{\D T} h_\sigma^{-1/2}&\quad
 x_q  :=  \frac{k}{T} h_q^{-1/2}\nonumber
 \ea
we obtain the flow equation e.g.~for the free energy density:
\ba\label{freeeq} 
 k \frac{\partial v_k (T)}{\partial k} & = &
 \frac{k^4}{2(4\pi)^2} \left[ 3 h_\pi + h_\sigma -8N_c h_q 
 + \frac{2k}{T} \Big[ 3 B_1 ( x_\pi) 
 + B_1 (x_\sigma)
 - 8N_c F_1 (x_q)\Big] \right]\nonumber\\
 \ea
Here one sees the separation of quantum and thermal fluctuations: 
The right hand 
side of the flow equation splits into the zero temperature flow 
equation part and a finite temperature part. 

\end{appendix}

\end{document}